\documentclass[12pt]{article}
\usepackage{jcappub}
\usepackage{adjustbox,multirow, longtable}
\usepackage{cancel,enumitem,dsfont,extarrows}
\usepackage{mathtools,amssymb,amsthm,amsmath,amsfonts}
\usepackage{xcolor,graphicx,subfigure}
\usepackage[utf8]{inputenc}
\usepackage[normalem]{ulem}
\usepackage{caption}
\usepackage{subcaption}
\allowdisplaybreaks

\newcommand{\ignore}[1]{}
\newcommand{\bk}{\boldsymbol{k}}
\newcommand{\p}{\boldsymbol{p}}
\newcommand{\bphi}{\boldsymbol{\phi}}
\newcommand{\bvarphi}{\boldsymbol{\varphi}}
\newcommand{\deltain}{\delta_{\rm in}}
\definecolor{darkgreen}{RGB}{50,150,0}

\arxivnumber{2505.10915}
\begin{document}
\title{Precision calculation of the EFT likelihood with primordial non-Gaussianities}
\author[a]{Ji-Yuan Ke,}
\author[a]{Yun Wang}
\author[a,b]{and Ping He}
\affiliation[a]{Center for Theoretical Physics and College of Physics, Jilin University, Changchun 130012, China}
\affiliation[b]{Center for High Energy Physics, Peking University, Beijing 100871, China}
\emailAdd{kejy22@mails.jlu.edu.cn, yunw@jlu.edu.cn, hep@jlu.edu.cn}

\abstract
{We perform a precision calculation of the effective field theory (EFT) conditional likelihood for large-scale structure (LSS) using the saddle-point expansion method in the presence of primordial non-Gaussianities (PNG). The precision is manifested at two levels: one corresponding to the consideration of higher-order noise terms, and the other to the inclusion of contributions from the saddle-point fluctuations. When computing the latter, we encounter the same issue of the negative modes as in the context of false vacuum decay, which necessitates deforming the original integration contour into a combination of the steepest descent contours to ensure a convergent and real result. We demonstrate through detailed calculations that, upon incorporating leading-order PNG, both types of extensions introduce irreducible field-dependent contributions to the conditional likelihood. This insight motivates the systematic inclusion of additional effective terms within the forward modeling framework. Our work facilitates Bayesian forward modeling under non-Gaussian initial conditions, thereby enabling more stringent constraints on the parameters describing PNG.}

\maketitle

\section{Introduction}

Primordial non-Gaussianities (PNG) are a pivotal and enduring area of interest in contemporary cosmology. On the one hand, they constitute a fundamental prediction of inflationary theory \cite{Guth:1980zm, Achucarro:2022qrl, Linde:1981mu, Guth:1982ec, Maldacena:2002vr, Cheung:2007st}, potentially providing a crucial observational window into the relics of the early universe (e.g. relics from physics beyond the standard model \cite{Lee:2016vti,Arkani-Hamed:2015bza}). On the other hand, non-Gaussian initial conditions can also impact the evolution of the dark matter and galaxy bias \cite{Dalal:2007cu, Giannantonio:2009ak, Tellarini:2015faa, Raccanelli:2015oma, Assassi:2015fma}, consequently leaving characteristic imprints on the large-scale structure (LSS) of the universe. Recent studies have been endeavoring to capture traces of PNG at large scales from both theoretical \cite{Wang:2024rdf, Uhlemann:2017tex, Friedrich:2019byw} and experimental \cite{DESI:2025qqy, Chaussidon:2024sxv, DESI:2024wki} perspectives.

This motivates the continuous development of more advanced methodologies to constrain the parameters describing PNG, among which PNG Bayesian forward modeling \cite{Andrews:2022nvv, Euclid:2024ris} is particularly prominent, as it optimally exploits all available information without the need of summary statistics. In the spirit of this approach, one should first account for various cosmological effects under specified initial conditions, then directly predict the present-day galaxy distribution, and finally compare the results with observational data until convergent. Within this framework, a key quantity requiring a semi-analytical treatment is the conditional likelihood $\mathcal{P}[\delta_g | \delta]$, which is the functional probability of observing a galaxy field $\delta_g$ given a realization of the matter field $\delta$. In the standard approach, one typically uses the properties of conditional probability to express this as a ratio of two terms: the matter likelihood $\mathcal{P}[\delta]$ and the joint likelihood $\mathcal{P}[\delta,\delta_g]$.

To date, considerable effort has been devoted to deriving analytic expressions for the two likelihoods based on perturbation theory \cite{Schmidt:2018bkr, Cabass:2019lqx, Ke:2025uou, Voivodic:2025quw, Cabass:2020nwf, Schmidt:2020tao}.  In this framework, the effective field theory of large-scale structure  (EFTofLSS) \cite{Baumann:2010tm, Carrasco:2012cv, Carroll:2013oxa, Porto:2013qua, Carrasco:2013mua, Konstandin:2019bay, Senatore:2014via} can be incorporated naturally into the likelihood construction. For instance, Schmidt et al. \cite{Schmidt:2018bkr} integrate out small-scale degrees of freedom, coarse-grain the underlying dynamics, and thereby obtain an EFT likelihood that governs the statistical properties on large scales. However, when calculations are based on standard perturbation theory (SPT) \cite{Bernardeau:2001qr}, one will inevitably omit higher-order contributions to the likelihood, as the evolution equation of $n$-point correlation functions involves higher-order information.\footnote{This can be found by comparing the results of \cite{Schmidt:2018bkr} and \cite{Cabass:2019lqx}. See \cite{Ke:2025uou} for a detailed discussion of the underlying reasons.} Consequently, we favor the recently developed functional approach \cite{Cabass:2019lqx, Ke:2025uou}, which solves this issue. Its origins can be traced to the time-sliced perturbation theory (TSPT) \cite{Blas:2015qsi, Blas:2016sfa, Vasudevan:2019ewf}, where it was shown that TSPT reproduces results from SPT without requiring higher-order information. For likelihood calculations, this method yields the likelihood directly by performing a functional Fourier transform of the LSS partition function, reducing the problem to a path integral in three-dimensional Euclidean space \cite{Ke:2025uou}. In this way, a wide range of cosmological effects can be incorporated phenomenologically by mapping them to distinct terms in the partition function, leading to a likelihood expression that consistently accounts for these effects.\footnote{In fact, this advantage of the functional approach extends well beyond likelihood computations. Several studies have employed the ``effective likelihoods'' as foundational inputs for deriving Feynman rules \cite{Vasudevan:2019ewf}, renormalization group equations \cite{Rubira:2024tea, Nikolis:2024kbx}, and other components of the theoretical framework.}

In our previous work \cite{Ke:2025uou}, we extended this functional approach beyond the saddle-point approximation. By adapting the saddle-point expansion technique, which was developed in quantum field theory to compute tunneling rates \cite{Coleman:1977py, Callan:1977pt}, we evaluated the additional contributions arising from fluctuations around the saddle points. Building on these results, the present study carries out a precision calculation of the EFT conditional likelihood in the presence of PNG. The standard approach to incorporating PNG \cite{Schmidt:2010gw, Assassi:2015jqa} introduces an effective vertex cubic in the initial matter field within the partition functions \cite{Nikolis:2024kbx}. In \cite{Cabass:2019lqx}, the authors investigated the case in which only the leading-order noise contributions are included, and argued that under this approximation, the corrections from PNG to both the matter likelihood and the joint likelihood are identical, thereby yielding a conditional likelihood that is consistent with that obtained under Gaussian initial conditions. Nevertheless, this work establishes that once more precision calculations are taken into account, nontrivial corrections to the conditional likelihood inevitably emerge.

We will demonstrate through explicit calculations that these nontrivial contributions primarily originate from two distinct sources. The first pertains to the incorporation of higher-order noise terms, which are systematically introduced to cancel the UV dependence of the loop diagrams. As such, their incorporation can be viewed as analogous to the renormalized effective action in quantum field theory \cite{Schwartz:2014sze}. The second originates from the incorporation of small fluctuations around the saddle points, i.e. applying the saddle-point expansion method. When calculating this contribution, it is essential to carefully examine whether the eigenvalues of the second derivative of the classical action are negative \cite{Coleman:1977py, Ai:2019fri, Andreassen:2016cvx}. Should this be the case, the steepest descent method in the Picard-Lefschetz theory \cite{Witten:2010cx} needs to be employed to approximate the integral, guaranteeing that the path integral remains convergent and real. In this paper, we will provide a concise review of this method and subsequently apply it to our computations. In general, the contributions from both sources are typically of next-order magnitude compared to the saddle-point contributions. However, in pursuit of more accurate and robust theoretical predictions, it is imperative to account for both classes of contributions within forward modeling. We anticipate that this work may offer valuable insights for Bayesian forward modeling under non-Gaussian initial conditions, thereby facilitating more stringent constraints on the parameters describing PNG. 

The outline of this paper is as follows. In section~\ref{sec:review}, we review the application of the saddle-point expansion method for calculating the EFT likelihood with a concrete example. Section~\ref{sec:addpng} is devoted to incorporating non-Gaussian statistics into the path integral formulation to the LSS. Subsequently, we present the explicit procedure for computing the conditional likelihood in section~\ref{sec:eftlikelihood}. Finally, we summarize our main findings and draw our conclusions in section~\ref{sec:conclusion}.

Throughout this paper, the notations and conventions used in the path-integral approach to the LSS mainly come from \cite{Cabass:2019lqx}.

\section{Review of the saddle-point expansion method}
\label{sec:review}

For the sake of clarity and familiarity, in this section we present a brief review of the saddle-point expansion method for computing the EFT likelihood, illustrated via a concrete example. This method was initially proposed in quantum mechanics and quantum field theory for the calculation of decay rates, and has led to reliable results that are consistent with the quantum mechanical estimates of tunneling rates \cite{Coleman:1977py, Callan:1977pt}. More recently, this method has been applied to the path-integral approach to the LSS, yielding more precision calculation of the conditional likelihood \cite{Ke:2025uou}. Specifically, we will focus on the origin of the negative (complex) modes and demonstrate how the original, potentially ill-defined integration contour can be systematically deformed into a sum of the steepest descent contours, thereby restoring convergence and physical consistency.

Suppose now we aim to calculate the joint likelihood of the following form, which is identical to that considered in \cite{Ke:2025uou, Cabass:2019lqx}:
\begin{equation}
      \mathcal{P}_g[\delta_g,\delta] = \int \mathcal{D} \bphi_g\,  e^{-S_g[\bphi_g]} \, , 
      \label{jointlikelihood}
\end{equation}
where $\bphi_g = (X_g , X, \deltain)^{T}$ represents the field that incorporates all the integral parameters, and $S_g[\bphi_g]$ is the ``effective action" of this theory, which in this example can be expressed as
\begin{equation}
      S_g[\bphi_g] = \int_{\bk} \bphi^{a}_{g}(\bk) \mathcal{J}^{a}_{g}(-\bk) + \frac{1}{2}\int_{\bk,\bk'} \mathcal{M}^{ab}_{g}(\bk,\bk') \bphi^{a}_{g}(\bk)\bphi^{b}_{g}(\bk')\, .  \label{eaction}
\end{equation}
The explicit formulae of the two matrices are given by $\mathcal{J}=(-i\delta,0)$ and
\begin{equation}
     {\cal M}_g(\bk,\bk') = (2\pi)^3\delta^{(3)}_{D}(\bk+\bk')
\begin{pmatrix}
P_{\epsilon_g}(k) & P_{\epsilon_g\epsilon_m}(k) & i K_{g,1}(k) D_1 \\
P_{\epsilon_g\epsilon_m}(k) & 0 & i K_1(k)D_1 \\
i K_{g,1}(k) D_1 & i K_1(k)D_1 & P^{-1}_{\rm in}(k)
\end{pmatrix}\,\,. \label{emformula}
\end{equation}
At zeroth order, we can simply take $K_1 = 1$ and $K_{g,1} = b_1$. It is evident that the computation of this likelihood is formally analogous to the evaluation of a multi-field path-integral in three-dimensional Euclidean space, which enables the application of the saddle-point expansion method in this context. The first step in implementing this method involves identifying the saddle points of this action, which satisfy the following equation
\begin{equation}
     \frac{\delta S_g}{\delta \bphi_g}\bigg|_{\bphi_g = \bar{\bphi}_g} = \mathcal{J}^{a}_{g} + \mathcal{M}^{ab}_{g}\bar{\bphi}_{g}^{b} =0 \,. \label{esaddleequation}
\end{equation}
In this example, there is only one saddle-point solution. The contribution from the saddle point will dominate the path integral, as it corresponds to the region where the action exhibits minimal oscillatory behavior. Therefore, we can decompose the field around the saddle point as $\bphi_g = \bar{\bphi}_g + \bvarphi_g$, where $\bvarphi_g$ represents the higher-order fluctuation terms,\footnote{In the context of quantum field theory, the field expansion typically takes the form $\phi = \bar{\phi} + \hbar^{1/2}\varphi$ (e.g. see \cite{Garbrecht:2015oea}). This allows the expansion term to be systematically treated as a higher-order variable. Moreover, it has been demonstrated in \cite{Ke:2025uou} that an analogous result holds at large scales of the universe, thereby justifying the applicability of this decomposition in our analysis.} which are also referred to as quantum fluctuations in quantum field theory. Substituting this decomposition into eq.~\eqref{jointlikelihood}, we can obtain
\begin{equation}
    \mathcal{P}_g[\delta_g,\delta] = \int \mathcal{D} \bphi_g \, {\rm}e^{-S_g[\bphi_g]}
\approx \int \mathcal{D} \bvarphi_g \, e^{-S_g[\bar{\bphi}_g] - \frac{1}{2}S_g''[\bar{\bphi}_g]\bvarphi_g^2 +\cdots}\,. \label{elikelihood}
\end{equation}
It can be seen that the saddle-point contribution $S_g[\bar{\bphi}_g]$ is independent of the remaining integration and may therefore be extracted as an overall prefactor. For the fluctuations around the saddle point, one must compute the eigenvalues of the second derivative of the action at the saddle point and then apply the Gaussian integral formula to evaluate their contributions. However, a complication arises when negative (complex) eigenvalues of $S''_g[\bar{\bphi}_g]$ appear in the calculation, rendering the entire integral ill-defined, i.e. leading to unphysical imaginary results \cite{Callan:1977pt}. For instance, from the expression for $\mathcal{M}_g$ in eq.~\eqref{emformula}, we can immediately observe that it is a general three-dimensional matrix. Consequently, its eigenvalues may involve negative or imaginary parts (we can examine this through direct calculations). Incorporating these eigenvalues into our framework would result in a complex likelihood, which is incompatible with its physical interpretation as a probability. 

In fact, the presence of negative modes is not an isolated case, but rather a longstanding issue that has persisted in the computation of tunneling rates. In the context of false vacuum decay, the predecessors ultimately identified the steepest descent method within the framework of Picard-Lefschetz theory as a solution to this problem, as illustrated in \cite{Callan:1977pt, Ai:2019fri, Andreassen:2016cvx}. This approach has more recently been demonstrated to be directly applicable to the computation of the likelihoods, yielding robust and reliable results \cite{Ke:2025uou}. Following the insights developed in earlier studies, we summarize the overall strategy in the following steps:

\textit{i) We complexify the action and identify all the complex saddle points.}

This implies that we extend the field $\bphi_g$ to the complex plane and perform the path integral on a middle-dimensional contour \cite{Ai:2019fri}. In our example, as seen in eq.~\eqref{esaddleequation}, this complexification does not introduce any new saddle-point solutions. Therefore, we can retain the previous solution. 

\textit{ii) For each saddle point in the complex plane, we determine a corresponding steepest descent contour.}

Intuitively, each steepest descent contour (also referred to as the Lefschetz thimble \cite{Witten:2010cx}) is defined simply by moving away from the corresponding saddle point in the direction that increases the real part of $S_g$ as quickly as possible. In this case, since the expression for the action involves multiple fields, it is highly nontrivial to identify these contours by direct inspection. Therefore, we resort to employing the gradient flow equation to determine these contours.

To achieve this, we parameterize the family of contours passing through the saddle point $\bar{\bphi}_g$ via $u$, and impose the boundary condition $\bphi_g(\boldsymbol{k},u \rightarrow -\infty) = \bar{\bphi}_g$. This implies that in the limit of $u\rightarrow -\infty$, we recover the saddle-point solution. To ensure the steepest descent of the real part of $S_g$, we choose the holomorphic function as $\mathcal{I}[\bphi_g] = - S_g[\bphi_g]$, and then define the Morse function $f[\bphi_g] = {\rm Re}(\mathcal{I}[\bphi_g])$. Thus, the gradient flow equation for $\bar{\bphi}_g$ is \cite{Witten:2010cx}
\begin{equation}
    \frac{\partial \bphi_g(\bk,u)}{\partial u} = - \overline{\left( 
    \frac{\delta\mathcal{I}[\bphi_g(\bk,u)]}{\delta\bphi_g(\bk,u)}\right)}, \ \  \ \ { \frac{\partial \overline{\bphi_g(\bk,u)}}{\partial u} }= - \left( 
    \frac{\delta\mathcal{I}[\bphi_g(\bk,u)]}{\delta\bphi_g(\bk,u)}\right). \label{gradient-flow-equation}
\end{equation}
Note that the overline here does not denote a saddle point, but rather the complex conjecture (following the convention adopted in \cite{Ai:2019fri}). By examining the non-positivity of $\partial f/ \partial u$, we can confirm that the condition imposed by the equation indeed selects the desired steepest descent contour. After substituting the expression for the action from eq.~\eqref{eaction} and employing the classical equation of motion from eq.~\eqref{esaddleequation}, we arrive at
\begin{equation}
\frac{\partial\bvarphi_g(\bk,u)}{\partial u} = \mathcal{M}^{*}_{g} \bvarphi^{*}_{g} \, . \label{egradient}
\end{equation}
The kernel of the steepest descent method lies in approximating the original integration contour by a combination of steepest descent contours that is homologous to it,\footnote{This procedure is generally feasible, as all convergent integration contours within this manifold -- and their equivalence relations -- form a homology group. See \cite{Ke:2025uou} for details.} thereby ensuring a convergent and real result. In this way, the calculation of the likelihood is reformulated as evaluating the path integral over all contours in the complex plane that are constrained by the given equation eq.~\eqref{egradient}.

\textit{iii) Following the above spirit, each Lefschetz thimble terminates at the convergent regions of the integral or at another saddle point. Therefore, each thimble provides a convergent path integral along the complex integration contour in field space.}

This conclusion is guaranteed by the Picard-Lefschetz theory \cite{Andreassen:2016cvx, Witten:2010cx} and can be explicitly verified within our example. A separation of variables approach can be applied to eq.~\eqref{egradient} by adopting the following ansatz: $\bphi_g(\bk,u) = \sum_n g_{n}(u)\chi_{n}(\bk)$, where $g_{n}(u) \in \mathbb{R}$ and the subscript ``$n$" distinguishes different directions. After implementing this ansatz, we obtain an eigen-equation involving $g_n(u)$ and $\chi_n(\bk)$
\begin{equation}
    \mathcal{M}_{g}^{*} \, \chi_{n}^{*}(\bk)/\chi_{n}(\bk) = m_{n} =\frac{1}{g_{n}(u)} \frac{{\rm d} g_{n}(u)}{{\rm d} u}\, , 
    \label{full-eigenequation}
\end{equation}
where we have introduced the real eigenvalues $m_n$, since the right-hand side (r.h.s) of this equation is manifestly real. Our task now reduces to solving the corresponding eigen-equation
\begin{equation}
     \mathcal{M}_{g}^{*} \, \chi_{n}^{*}(\bk) = m_{n} \,\chi_{n}(\bk) \,. \label{eigenequation}
\end{equation}
From the explicit form of $\mathcal{M}_g$ in eq.~\eqref{emformula}, it is evident that $\mathcal{M}_{g}^{*}$ is a general three-dimensional matrix, which renders the eigenvalue problem analytically challenging. Moreover, since $\mathcal{M}_{g}^{*}$ is neither Hermitian nor anti-Hermitian, we cannot generally expect its eigenstates to be purely real or imaginary. To address this issue, we consider this eigen-equation eq.~\eqref{eigenequation} together with its complex conjecture $\mathcal{M}_{g} \, \chi_{n}(\bk) = m_{n} \,\chi_{n}^{*}(\bk)$, and construct from them a Hermitian eigen-equation
\begin{equation}
        \begin{pmatrix}
         \boldsymbol{0}  & \mathcal{M}^{*}_{g} \\
         \mathcal{M}_{g} & \boldsymbol{0}
        \end{pmatrix}
    \begin{pmatrix}
        \chi_{n}(\bk) \\ {\chi}_{n}^{*}(\bk)
    \end{pmatrix}
     =  m_{n}\begin{pmatrix}
         \chi_{n}(\bk) \\ {\chi}_{n}^{*}(\bk)
     \end{pmatrix} \, . \label{hermite-equation}
\end{equation}
This equation is now amenable to solution, with all eigenvalues $m_n$ being real by construction. On the other hand, from the eigen-equation on the r.h.s of eq.~\eqref{full-eigenequation}, it is evident that the solution for $g_{n}(u)$ takes the form  $g_{n} (u) \sim a_{n}\,\exp\,(m_nu)$, where $a_n \in \mathbb{R}$. Under the boundary condition we have imposed, $g_n(u\rightarrow - \infty) =0$, it can be shown that all eigenvalues $m_n$ are positive and real. This ensures both the convergence and reality of the integral along the Lefschetz thimbles.

\textit{iv) With these thimbles in field space, we can approximate the integration contour as a combination of different Lefschetz thimbles that homology to the original integral contour to obtain a convergent and real result.}

The eigenvalues of eq.~\eqref{hermite-equation} have already been solved in \cite{Ke:2025uou}, and we omit the corresponding algebra and adopt their results in this work: 
\begin{equation}
    \sum_{n} m_n =m_1+m_2+m_3 = |P_{\epsilon_g} - P_{\rm in}^{-1}| \,. \label{sum-of-mn}
\end{equation}
Here, we adopt a more rigorous formulation.\footnote{In fact, eq.~\eqref{hermite-equation} admits six eigenvalues, but only three are of interest in our context. Due to the algebraic approximation taken in the calculation, the result of $\sum_n m_n$ could be either $P_{\epsilon_g} -  P_{\rm in}^{-1}$ or $- P_{\epsilon_g} +  P_{\rm in}^{-1}$ (we need to select the non-negative solution). Consequently, we express the result in terms of its absolute value which is non-negative.} Plugging this into eq.~\eqref{elikelihood} allows us to compute the likelihood
\begin{align}
    \mathcal{P}_g[\delta_g,\delta] & \approx  \int_{C} \mathcal{D} \boldsymbol{\phi}_g \,  e^{-S_g[\bar{\bphi}_g] -\frac{1}{2} \bvarphi_g S''_g[\bar{\bphi}_g]\bvarphi_g}  = \int_C \mathcal{D} \bvarphi_g \,  e^{-S_g[\bar{\bphi}_g] - \bvarphi_g M_g\bvarphi_g}\nonumber\\ 
    & =\int \mathcal{D} \bvarphi_g \,e^{-S_g[\bar{\bphi}_g] - \sum_n m_n|\bvarphi_g|^2} =  e^{-S_g[\bar{\bphi}_g]} \frac{\pi}{\sum_n m_n} \approx e^{-S_g[\bar{\bphi}_g]} \frac{\pi}{|P_{\epsilon_g}-P_{\rm in}^{-1}|}.   \label{efinal-result}
\end{align}
Let us provide a detailed account of the steps involved in the calculation. In the first line, we first approximate the integration contour as the combination of different Lefschetz thimbles $C$, in accordance with the implementation of the steepest descent method. Then we substitute the expression for $S''_g[\bar{\bphi}_g]$. In the second line, we first employ the gradient flow equation to constrain the integration contour to be composed of all steepest descent contours in the complex plane. Finally, we make use of eq.~\eqref{sum-of-mn}. 

In summary, the saddle-point expansion method approximates the full path integral by incorporating contributions from the saddle points and their neighborhoods, which may, in certain cases, yields complex results. However, when employing the steepest descent method, although the integrals along individual Lefschetz thimbles may lead to imaginary contributions, these imaginary parts are exactly canceled by those from other thimbles, resulting in a manifestly real total expression. The cancellation of these imaginary components and the guarantee of a real result constitute the core spirit of the application of the Picard-Lefschetz theory. This method has demonstrated remarkable efficacy in calculating the non-perturbative effects of quantum tunneling, facilitating the estimation of the temporal occurrence of a first-order phase transition in the early universe (see, for example, \cite{Quiros:1999jp, Moreno:1998bq} for details and \cite{Ke:2024lel} for applications). We also anticipate that this method could prove equally effective in calculating other Euclidean-space path integrals, thereby aiding in more precision calculations of the corresponding physical quantities. In fact, the use of the term ``precision" in the title of this paper is a deliberate inheritance from the terminology employed in \cite{Andreassen:2016cvx} to characterize the computation of tunneling rates. Moreover, the existence of negative modes does not indicate a flaw in the theory, but rather a consequence of incorrectly applying the steepest descent method, as illustrated in \cite{Ai:2019fri}.

\section{Generalizing the likelihoods to the non-Gaussian regime}
\label{sec:addpng}

One significant advantage of the path-integral approach to the LSS is that different cosmological effects can be systematically encoded as corresponding effective terms in the partition functions, in accordance with the spirit of Effective Field Theory \cite{Baumann:2010tm, Carrasco:2012cv, Senatore:2014via}. The objective of this section is to introduce PNG and elucidate how their effects can be incorporated into the partition functions. As we shall see, at leading order, the implementation of PNG is equivalent to introducing a cubic interaction term in the initial matter field within the partition functions. This term is proportional to both $f_{\rm NL}$ and $\deltain^3$, which requires careful treatment in the computation of the likelihoods. At the same time, this feature allows us to employ perturbative methods, as in section~\ref{sec:eftlikelihood}, to compute the leading-order $f_{\rm NL}$ corrections to the likelihoods.

\subsection{Introducing primordial non-Gaussianities}
We begin with the standard formulation for incorporating PNG, where the core idea is to expand the primordial non-Gaussian potential as a series around the Gaussian potential \cite{Schmidt:2010gw, Assassi:2015fma}
\begin{equation}
    \varphi_{\rm NG}(\bk) = \varphi(\bk) + f_{\rm NL} \int_{\p} K_{\rm NL}(\p,\bk-\p)\left[\varphi(\p) \varphi(\bk-\p)-P_{\rm G}(p)\hat{\delta}_{D}(\bk)\right] + \cdots\, , \label{primordialpotential}
\end{equation}
where $\varphi$ is a Gaussian random variable, $P_{\rm G}(k) \equiv \left \langle \varphi(\bk) \varphi(-\bk) \right \rangle'$,  in which $\left<\cdots\right>'$ indicates that an overall momentum-conserving delta function has been dropped. We have also defined $\hat{\delta}_D(\bk_1 +\bk_2) = (2\pi)^3 \delta_D(\bk_1 + \bk_2)$. The $\cdots$ in eq.~\eqref{primordialpotential} represents higher-order non-Gaussian contributions. Throughout this paper, we assume that the fields without the subscript ``NG" correspond to Gaussian fields and only consider the leading-order non-Gaussianities. $K_{\rm NL}$ represents the kernel function, which is dependent on the inflationary models. This expression leads to a non-vanishing primordial bispectrum
\begin{align}
    B_{\varphi}(k_1,k_2,k_3) &\equiv \left\langle \varphi_{\rm NG}(\bk_1) \varphi_{\rm NG}(\bk_2) \varphi_{\rm NG}(\bk_3) \right\rangle' \nonumber \\ &= 2f_{\rm NL} K_{\rm NL} (\bk_1,\bk_2) P_{\varphi}(k_1)P_{\varphi}(k_2) + 2 \;\text{perms}\,,
\end{align}
where ``perms" denotes the cyclic permutations of the three momenta $k_1$, $k_2$, and $k_3$. In the squeezed limit, which is of critical importance for the study of galaxy clustering, the non-singular kernels can be uniquely determined \cite{Assassi:2015fma}.\footnote{It is worth noting that the existence of particles beyond the standard model with masses comparable to the Hubble scale during inflation can potentially induce non-analytic features in the squeezed limit through the cosmological collider mechanism \cite{Chen:2009zp, Arkani-Hamed:2015bza}.}  In this case, we have 
\begin{equation}
    \frac{B_{\varphi}(k_{\ell}, |\bk_{s}-\frac{1}{2}\bk_{\ell}|, |\bk_{s}+\frac{1}{2}\bk_{\ell}|)}{P_{\varphi}(k_{\ell})P_{\varphi}(k_s)} \xrightarrow{k_{\ell} \ll k_s} 2f_{\rm NL}\left[ K_{\rm NL}(\bk_{\ell},\bk_s) + K_{\rm NL}(\bk_{\ell},-\bk_s) \right]\,.
\end{equation}
Under the assumption of spatial homogeneity and isotropy, the kernel $K_{\rm NL}(\bk_{\ell},\bk_{s})$ can only depend on the magnitudes of the two momenta $k_{\ell}$, $k_s$, as well as their relative angle $\hat{\bk}_{\ell} \cdot \hat{\bk}_s$. Furthermore, due to the $\bk_s \rightarrow -\bk_s$ symmetry of the bispectrum that arises in the squeezed limit, the kernel function can be expanded in a series of even-order Legendre polynomials \cite{Assassi:2015jqa, Assassi:2015fma}
\begin{equation}
    K_{\rm NL}(\bk_{\ell}, \bk_s) \xrightarrow{k_{\ell} \ll k_s} \sum_{L,i}a_{L,i} \left(\frac{k_\ell}{k_s}\right) ^{\Delta_{i}} \mathcal{P}_{L}(\hat{\bk}_{\ell} \cdot \hat{\bk}_s)\,,
\end{equation}
where $\mathcal{P}_{L}$ is the Legendre polynomial of even order $L$. This expression can encompass the predictions of various inflationary models \cite{Cheung:2007st, Komatsu:2001rj, Alishahiha:2004eh, Chen:2009zp, Green:2013rd, Arkani-Hamed:2015bza, Lee:2016vti, Flauger:2016idt}. In some literature, $L$ is commonly referred to as the spin of the non-Gaussianity, with different values leading to distinct additional effective operators in the bias expansion \cite{Assassi:2015fma, Nikolis:2024kbx}. These operators also play a crucial role in the study of PNG, for instance, within the galaxy bias renormalization group approach \cite{Rubira:2023vzw, Rubira:2024tea, Nikolis:2024kbx}, they induce nontrivial contributions to the running of the bias parameters, thereby affecting the predictive structure of LSS observables. We note that, in order to incorporate the contributions from these operators, we are supposed to modify the term related to the bias expansion, $\delta_{g, \rm fwd}[\deltain]$ (see the next section), within the galaxy partition function. However, we emphasize that the computation of the likelihoods in this context does not rely on the specific form of the bias expansion, so we can appropriately omit these additional expansion operators and generalize them when required. In the subsequent calculations, we will consider only the leading-order bias expansion as an illustrative example, i.e. $\delta_g = b_1\delta$.

\subsection{The partition functions with primordial non-Gaussianities}

We now proceed to construct the partition functions of LSS in the presence of PNG. To achieve this, we need to establish a connection between the primordial potential and the initial matter field. We first define the transfer function $T(k)$ to relate $\varphi(\bk)$ to the linearly-evolved potential
\begin{equation}
    \Phi_{(1)} (\bk) = T(k) \,\varphi(\bk) \, .
\end{equation}
In this expression, we have omitted the dependence of the transfer function $T$ on the conformal time $\tau$, which should be kept in mind. The normalization condition of this function is $T(k) \rightarrow 1$ for $k\ll k_{\rm eq}$. We then employ the gravitational Poisson equation in Fourier space \cite{Schmidt:2010gw}
\begin{equation}
    \deltain (\bk) = \frac{2}{3} \frac{k^{2} D(\tau)}{H_{0}^{2} \Omega_{m,0}} \Phi_{(1)} (\bk),
\end{equation}
where $H_0$ is the present-day Hubble constant, $\Omega_{m,0}$ is the fractional matter density, and $D$ is the growth factor. From these two expressions, we obtain the relation between the primordial potential and the initial matter field
\begin{equation}
    \deltain(\bk) = M(k,\tau) \, \varphi(\bk), \, \: \; M(k,\tau) = \frac{2}{3}\frac{k^2 T(k) D(\tau)}{H_{0}^{2} \Omega_{m,0}}\,. \label{pngdensity}
\end{equation}
Analogously, we define the non-Gaussian matter field $\delta_{\rm in, NG}$ such that eq.~\eqref{pngdensity} naturally induces its expansion into the following form
\begin{equation}
    \delta_{\rm in, NG} (\bk) = \deltain (\bk) + f_{\rm NL} \int_{\p_1,\p_2} \hat{\delta}_{D}(\bk-\p_{12}) K_{\rm NL}(\p_1,\p_2) 
\frac{M(|\p_1 +\p_{2}|)}{M(p_1)M(p_2)} \deltain(\p_1)\deltain(\p_2)\,. \label{pngmatter}
\end{equation}
 The contributions from higher-order PNG can be obtained through an analogous procedure. This expression allows for a straightforward incorporation of arbitrary-order PNG into the partition functions via appropriate corrections to the initial matter field $\deltain(\bk)$. For example, when we want to generalize the matter partition function presented in \cite{Cabass:2019lqx}
\begin{equation}
    Z[J] = \int \mathcal{D}\deltain \, \exp \left\{ \int_{\bk} \left(\frac{1}{2}P_{\epsilon_{m}}(k)J(\bk)J(-\bk) +J(\bk)\delta_{\rm fwd}[\deltain](-\bk)\right) \right\}\mathcal{P}[\deltain]\,, 
\end{equation}
where $\mathcal{P}[\deltain]$ is the initial probability distribution, which is usually assumed to be Gaussian, and  $P_{\epsilon_m}(k)$ denotes the power spectrum of the noise in the matter density field, which is used to cancel the UV dependence of the loop diagrams \cite{Cabass:2019lqx}. At this point, all that is required is to substitute the non-Gaussian statistics of $\deltain$ eq.~\eqref{pngmatter} into the probability distribution  
\begin{align}
    \ln \mathcal{P}[\deltain] &= -\frac{1}{2} \int_{\bk} \frac{\deltain(\bk)\deltain(-\bk)}{P_{\rm in}(k)} \rightarrow -\frac{1}{2} \int_{\bk} \frac{\deltain(\bk)\deltain(-\bk)}{P_{\rm in}(k)} \nonumber \\
    &-f_{\rm NL}\int_{\p_1,\p_2,\p_3}\left 
    [\hat{\delta}_{D}(\p_{123}) K_{\rm NL}(\p_2,\p_3) \frac{M(p_1)}{M(p_2)M(p_3)}\frac{\deltain(\p_1)}{P_{\rm in}(p_1)} \deltain(\p_2)\deltain(\p_3) \right] \,. \label{addpng}
\end{align}
Under this adjustment, the original partition function is modified accordingly and takes the following form
\begin{align}
    Z[J] &= \int \mathcal{D} \deltain \, \exp \left\{\int_{\bk} \left(\frac{1}{2}P_{\epsilon_{m}}(k)J(\bk)J(-\bk) +J(\bk)\delta_{\rm fwd}[\deltain](-\bk)\right) \right\} \nonumber \\
    & \times \exp \left\{- \frac{1}{3!}\,f_{\rm NL} 
    \int_{\bk,\p_1,\p_2} \hat{\delta}_{D}(\bk +\p_{12}) B(\bk,\p_1,\p_2) \deltain(\bk)\deltain(\p_1) \deltain(\p_2) \right\} \mathcal{P}[\deltain]\, , \label{pngmatter-partitionfunction}
\end{align}
where we have defined
\begin{equation}
   \frac{1}{3!} B(\bk,\p_1,\p_2) = K_{\rm NL}(\p_1,\p_2) \frac{M(k)}{M(p_1)M(p_2)} \frac{1}{P_{\rm in}(k)}\, . \label{nonGaussian-factor}
\end{equation}
Note that the sign convention adopted here is consistent with \cite{Vasudevan:2019ewf}, but opposite to that in \cite{Nikolis:2024kbx}.\footnote{In fact, both sign conventions are equally valid and yield the correct $n$-point statistics, as explicitly shown in eq.~(4.2) of \cite{Vasudevan:2019ewf} and eq.~(B.5) of \cite{Nikolis:2024kbx}. The difference between them merely corresponds to a redefinition of the ``coupling parameter" $B(\bk,\p_1,\p_2)\rightarrow-B(\bk,\p_1,\p_2)$. After matching with the $n$-point correlation functions, this distinction does not affect any physical predictions.} In this context, the coupling parameter $B(\bk, \p_1,\p_2)$ does not exhibit symmetry under cyclic permutations of $k,p_1,p_2$, but instead retains symmetry only under the exchange of $p_1$ and $p_2$. At the same time, $\mathcal{P[\deltain]}$ is still modeled as a Gaussian probability distribution, despite the fact that the underlying primordial statistics are no longer Gaussian. It is evident that, after incorporating PNG, the leading-order correction in $f_{\rm NL}$ effectively manifests as an additional interaction term in the partition function. Moreover, as illustrated in \cite{Nikolis:2024kbx}, this term can be absorbed into the term proportional to $J$ after performing a shift of the integration variable, which is consistent with the arguments in quantum field theory. Following the same procedure, we can derive the expression for the galaxy partition function with PNG
\begin{align}
    Z[J_g,J] &= \int \mathcal{D} \deltain \, \exp\left\{\int_{\bk}\left[\frac{1}{2}P_{\epsilon_g}(k)J_g(\bk)J_g(-\bk) + P_{\epsilon_g\epsilon_m}(k)J_g(\bk)J(-\bk) \right]\right\} \nonumber \\ 
     & \times \exp\left\{\int_{\bk}\left[\frac{1}{2}P_{\epsilon_m}(k)J(\bk)J(-\bk) +J_g(\bk)\delta_{\rm g,fwd}[\deltain](-\bk) +J(\bk)\delta_{\rm fwd}[\deltain](-\bk) + \cdots  
     \right]\right\} \nonumber\\
     & \times \exp \left\{- \frac{1}{3!}\,f_{\rm NL} 
    \int_{\bk,\p_1,\p_2} \hat{\delta}_{D}(\bk +\p_{12}) B(\bk,\p_1,\p_2) \deltain(\bk)\deltain(\p_1) \deltain(\p_2) \right\} \mathcal{P}[\deltain]\, . \label{pngjoint-partitionfunction}
\end{align}
Compared to eq.~\eqref{pngmatter-partitionfunction}, we have not only extended the current to separately correspond to the matter and galaxy fields, but have also generalized the stochastic term. This results in (a) the stochasticity for galaxies $P_{\epsilon_g} \sim k^0$, (b) the cross stochasticity between galaxies and matter $P_{\epsilon_g \epsilon_m} \sim k^2$, and (c) the matter stochasticity $P_{\epsilon_m} \sim k^4$ (where $\sim$ represents their leading-order contributions). When evaluating the joint likelihood in the large-scale limit, the matter stochasticity can usually be safely omitted, since it involves higher-order momentum dependence, which allows us to take $P_{\epsilon_m} \rightarrow 0$ \cite{Cabass:2019lqx}. 

In summary, in both cases, the effects of PNG can equivalently be reformulated as an interaction term under Gaussian initial conditions, thereby enabling the direct application of all analyses developed within the Gaussian framework. For example, in the current context, the matter likelihood can still be expressed as the functional Fourier transform of the matter partition function
\begin{equation}
    \mathcal{P}[\delta] =\mathcal{N}_{\delta^{(\infty)}}\, \int \mathcal{D} X \, {\rm exp} \left\{i\int_{\boldsymbol{k}}X(\boldsymbol{k})\delta(-\boldsymbol{k})\right\} Z[-iX]\,, \label{pngmatter-likelihood}
\end{equation}
where $\mathcal{N}_{\delta^{(\infty)}}$ is the normalization factor introduced in the definition of the functional Dirac function, and we have rewritten the current corresponding to $\delta$ as $X = iJ$. By employing the same methodology, we can derive the expression for the joint likelihood
\begin{equation}
    \mathcal{P}[\delta_g,\delta] = \mathcal{N}_{\delta^{(\infty)}}^{2} \int\mathcal{D}X_g \mathcal{D}X \, {\rm exp}\left\{i\int_{\bk} \left[X_g(\bk) \delta_g(-\bk)+X(\bk)\delta(-\bk)\right]\right\} Z[-iX_g,-iX], 
    \label{pngjoint-likelihood}
\end{equation}
where $X_g = iJ_g$ represents the current for the galaxy field. As we shall see, the normalization factor $\mathcal{N}_{\delta^{(\infty)}}$ will absorb all field-independent terms in the likelihood results, as illustrated in \cite{Cabass:2019lqx}. In the subsequent calculations, we will omit this factor provisionally and recover it at the final stage.

Here, we provide a brief comparison between the path-integral formulation and the standard perturbation theory (SPT) approach (see, for example, in \cite{Bernardeau:2001qr}) in the context of incorporating PNG. In SPT, the introduction of PNG leaves the dark-matter equations of motion unchanged; instead, it modifies the initial conditions from which the perturbative expansion begins. At leading order, for example, one must specify a non-trivial initial bispectrum and include its contributions in the loop-level matter power spectra \cite{Vasudevan:2019ewf}. By contrast, in the path-integral formulation, although the underlying physics is the same, this framework admits a more economical, phenomenological treatment. As shown in eq.~\eqref{addpng}, PNG changes the initial probability distribution $\mathcal{P}[\deltain]$, but its effects can be absorbed into the partition functions as an effective interaction vertex (see eqs.~\eqref{pngmatter-partitionfunction} and \eqref{pngjoint-partitionfunction}). The initial distribution can therefore still be treated as Gaussian, while all PNG effects are encoded in a single additional vertex, greatly simplifying the likelihood calculation. Moreover, the coefficients of these vertices are fixed by requiring that the path-integral formulation reproduces the same statistical properties obtained from the equation-of-motion approach. In the next section we will implement this strategy, use the saddle-point expansion to compute both likelihoods, and show how PNG introduces new, non-trivial contributions to the conditional likelihood.

\subsection{Dimensional analysis}

As we can see, the expressions for the two likelihoods, eqs.~\eqref{pngmatter-likelihood} and \eqref{pngjoint-likelihood}, involve numerous terms to be handled. Evaluating these likelihoods directly with all terms included poses a considerable challenge. Therefore, in this subsection, we perform a dimensional analysis to investigate the relative contributions of each term to the likelihoods at large scales. This can be achieved by performing a momentum rescaling analogous to the procedure outlined in \cite{Cabass:2019lqx}, with $\boldsymbol{k} = b \boldsymbol{k}'$. Then, in the limit $b\rightarrow 0^{+}$, the behavior of different terms under this rescaling will reveal their relevance (or irrelevance) to the large-scale universe. Under this rescaling, the integration measure ${\rm d}^3 k$ changes to $b^3{\rm d}^3k'$, and the power spectrum $P_{\rm in}$ scales as $P_{\rm in}(k) = b^{n_{\delta}}P_{\rm in}(k')$, where $n_{\delta}$ is the order of momentum in the initial power spectrum. Then, the terms in the partition functions that are quadratic in $X_g$ and $\delta_{\rm in}$ remain invariant if we redefine
\begin{equation}
    X(b\bk') = b^{-\frac{3}{2}}X_{g}'(\bk') \,, \,
    X_g(b\bk') = b^{-\frac{3}{2}}X_{g}'(\bk') \,, \,
    \delta_{\rm in} (b\bk') = b^{\frac{n_\delta-3}{2}} \deltain (\bk')\,.
\end{equation}
These are the results of the rescaling in Fourier space, and the scaling behavior of these operators in real space can be obtained via the inverse Fourier transform
\begin{equation}
    X(\boldsymbol{x}) \sim b^{\frac{3}{2}} \, , \; X_g(\boldsymbol{x}) \sim b^{\frac{3}{2}}\, , \;
    \deltain(\boldsymbol{x}) \sim b^{\frac{3+ n_{\delta}}{2}} \,. \label{realspace-rescaling}
\end{equation}
By employing a similar argument, we can derive the rescaling relations for the matter field $\delta(\boldsymbol{x})$ and the galaxy field $\delta_{g}(\boldsymbol{x})$ \cite{Cabass:2019lqx}
\begin{equation}
    \delta(\boldsymbol{x}) \sim b^{\frac{3+n_\delta}{2}} \, , \; \delta_{g}(\boldsymbol{x}) - b_1\delta(\boldsymbol{x}) \sim b^{\frac{3}{2}} \, .
\end{equation}
To determine the rescaling of the non-Gaussian effective vertex, it is also necessary to study the scaling behavior of the non-Gaussian parameter $B(\bk,\p_1,\p_2)$ in eq.~\eqref{nonGaussian-factor}. We begin in Fourier space, given the asymptotic behavior of the transfer function $T(k) \rightarrow 1$ in the limit $k\rightarrow 0$, it can be treated as dimensionless with respect to momentum at large scales and accordingly exhibits a trivial scaling behavior, namely $b^{0}$. In this way, $M(k)$ can be approximated to have a scaling behavior of $k^2$ in the relevant regime. Meanwhile, for the lowest-order momentum-dependent PNG, corresponding to the spin-0 case discussed in \cite{Assassi:2015fma}, the associated kernel function can also be considered momentum dimensionless. After substituting the momentum dependence of the initial power spectrum, the rescaling behavior of this parameter in real space is given by
\begin{equation}
B(\boldsymbol{x},\boldsymbol{x}_1,\boldsymbol{x}_2) \, \sim \, b^{2+n_{\delta}} \xrightarrow{n_{\delta} \rightarrow -2} \,b^{0}\, .
\end{equation}
Therefore, it is well justified to neglect the contributions arising from the momentum dimension of this parameter.

\begin{table}
\centering
\begin{tabular}{|c|c|}
\hline
Term & Scaling dimension \\ \hline
$\deltain\deltain$ & $b^{3+n_{\delta}}$\\ \hline
$P_{\epsilon_g}J_gJ_g$& $b^3$  \\ \hline
$P_{\epsilon_g\epsilon_m}J_gJ$ & $b^{5}$ \\ \hline
$J_g\delta_{g, \rm fwd}$ & $b^{3}$ \\ \hline
$J\delta_{\rm fwd}$ & $b^{3+n_\delta/2}$ \\ \hline
$\deltain\deltain\deltain$ & $b^{3(3+n_\delta)/2}$ \\ \hline
\end{tabular}
\caption{The scaling of different terms in eq.~\eqref{pngjoint-partitionfunction} under the rescaling $\bk = b \bk'$.}
\label{tab:scaling}
\end{table}

\begin{figure}
    \centering    \includegraphics[width=0.7\linewidth]{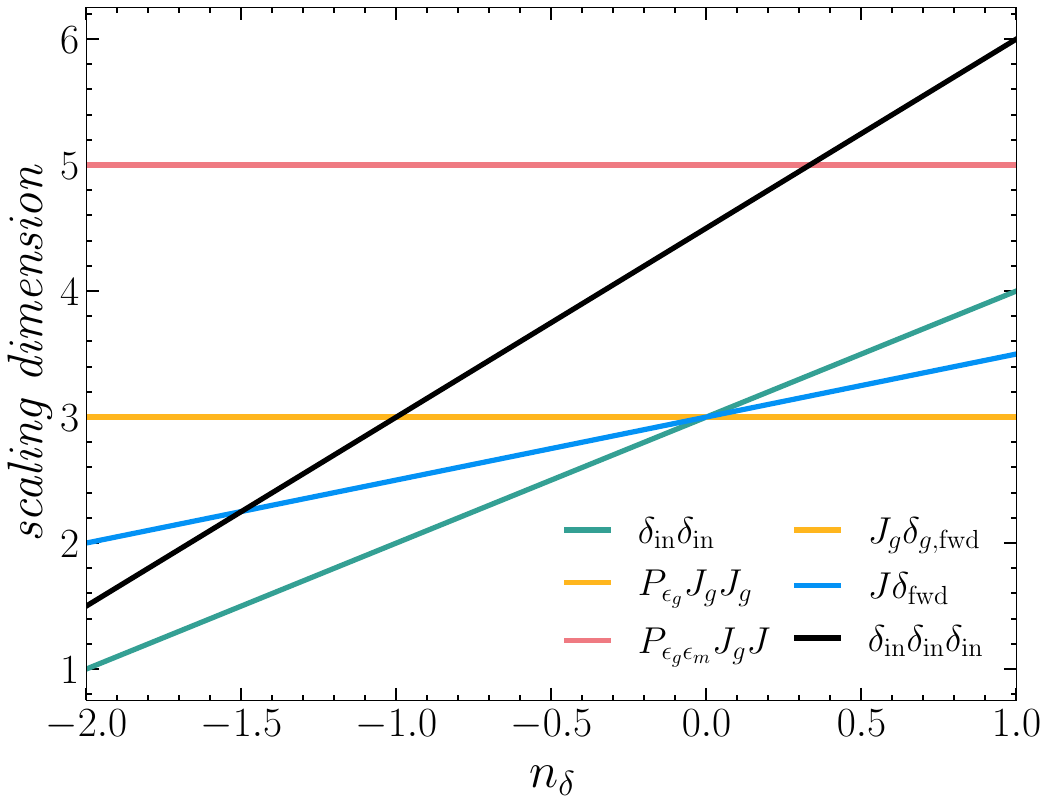}
    \caption{The evolution of the scaling dimensions of different terms in eq.~\eqref{pngjoint-partitionfunction} as a function of $n_\delta$.}
    \label{fig:scaling-dimension}
\end{figure}

Based on these expressions, we can assess the contributions of the different terms in the partition function eq.~\eqref{pngmatter-partitionfunction} at large scales. We present the result of the rescaling of these terms and the evolution of their scaling dimensions with respect to $n_\delta$ in Table~\ref{tab:scaling} and Figure~\ref{fig:scaling-dimension}, respectively. It is evident that for $n_\delta = -1.7$, both stochastic terms $P_{\epsilon_g}JJ$ and $P_{\epsilon_g\epsilon_m}J_gJ$ exhibit rescaling with a higher-order dependence on $b$ relative to the PNG effective term, thereby yielding subdominant contributions at large scales. This is precisely why, in \cite{Nikolis:2024kbx}, the authors were able to disregard the stochastic terms and focus primarily on the non-Gaussian contributions when deriving the renormalization group equations incorporating PNG. At the same time, we can observe that in our case,  the $P_{\epsilon_g\epsilon_m}J_gJ$ term generally contributes the least to the partition function. This is because the leading-order momentum dependence of $P_{\epsilon_g\epsilon_m}(k)$ scales as $k^2$, resulting in a stronger suppression compared to other operators. Consequently, it is reasonable to neglect the higher-order contributions from $P_{\epsilon_g\epsilon_m}$ when calculating the likelihoods. The perturbative properties of $f_{\rm NL}$ and $P_{\epsilon_g\epsilon_m}$ will constitute the central framework for our detailed calculations in the next section, enabling us to extract the leading-order additional terms associated with the non-Gaussian contributions.

In fact, higher-order terms such as $JJJ$ should also be included in the construction of the partition functions to ensure the theory's completeness under renormalization. However, in the present analysis, we neglect these terms for one primary reason: their contributions are negligible at large scales. To substantiate this claim, we refer to the dimensional analysis in \cite{Rubira:2024tea}, where the authors systematically investigate the dimensions of $J^n$ terms in real space. For a generic coupling between $J^n$ and an arbitrary operator $O[\delta]$, the effective vertex in the partition function and the dimension of its coefficient take the following form
\begin{equation}
    S_{\rm eff} \propto C_{O}^{(n)}(\Lambda) \int_{\boldsymbol{x}} [J(\boldsymbol{x})]^n O[\delta](\boldsymbol{x}) \,, \qquad \qquad [C_{O}^{(n)}] = 3-3n-d_{O} \, ,
\end{equation}
where $d_O$ is the dimension of the operator $O[\delta]$. As evident from this expression, the dimension of the coefficient decreases rapidly with increasing $J$-power. For instance, in the case of $JJJ$ term, this corresponds to $n=3$ and $d_O = 0$, such that $[C^{(3)}] = -6$. Consequently, the dimensionally consistent form is $\frac{C^{(3)}}{R^6}JJJ$, where $R$ represents the physical scale of interest (which is very large in our case). Therefore, higher-order terms are thus increasingly suppressed at large scales, justifying the neglect of terms beyond $J^3$ in likelihood computations.

As we have seen, two quantities in the partition functions eqs.~\eqref{pngjoint-partitionfunction} and \eqref{pngmatter-partitionfunction} are suppressed at large scales: the cross stochasticity term $P_{\epsilon_g\epsilon_m}$, suppressed due to its higher-derivative nature, and the PNG parameter $f_{\rm NL}$, suppressed by the small amplitude of PNG. We now provide a brief estimate of their magnitudes. We first consider the $P_{\epsilon_g\epsilon_m}$ term, for which the scale dependence can be factorized following the arguments in \cite{Cabass:2020nwf} (this term corresponds to the $P_{\epsilon_g}^{\{2\}}$ term in that literature),
\begin{equation}
    P_{\epsilon_g\epsilon_m}(k) \sim A R_{*}^{2} k^2 \, , \label{cross-scale}
\end{equation}
where $R_*$ is the typical nonlocal scale of galaxy formation, $A$ is some dimensionless constant and $k$ is the wavenumber corresponding to the scale in our study. In the context of our analysis, $R_*$ can be naturally identified with the nonlinear scale, $R_* \sim 1/k_{\rm NL}$. Substitute this into the expression for $P_{\epsilon_g\epsilon_m}$ in eq.~\eqref{cross-scale}, we can obtain
\begin{equation}
    P_{\epsilon_g\epsilon_m}(k) \sim A \left(\frac{k}{k_{\rm NL}} \right)^2 \sim \left(\frac{10h^{-1}\text{Mpc}}{1000 h^{-1}\text{Mpc}}\right)^2 \sim 10^{-4} \, ,
\end{equation}
where we have fixed the nonlinear scale to $10 h^{-1} \text{Mpc}$ and the scale of interest to $1000 h^{-1}\text{Mpc}$. It should be note that the magnitude of $P_{\epsilon_g\epsilon_m}$ is scale-dependent. On the other hand, the treatment of the non-Gaussian contributions is a little more subtle. Taking the local-type non-Gaussianity model as an example, based on the most recent constraints from the Planck satellite, $f_{\rm NL}^{\rm local} = -0.9 \pm 5.1, $ \cite{Planck:2019kim}, the parameter $f_{\rm NL}$ may be consistently taken to be of order $\mathcal{O}(1)$. Then this analysis indicates that the large-scale contribution of $P_{\epsilon_g\epsilon_m}$ is subdominant compared to that from $f_{\rm NL}$. It is therefore reasonable to neglect the higher-order terms in $P_{\epsilon_g\epsilon_m}$, as was done in \cite{Cabass:2019lqx}. However, current observations do not entirely exclude the possibility of a small but non-vanishing level of non-Gaussianity, as generally predicted by slow-roll inflationary scenarios \cite{Maldacena:2002vr}, e.g. $f_{\rm NL} \sim \mathcal{O}(10^{-2})$. Therefore, We do not rule out the possibility that the contribution of $f_{\rm NL}$ may be subdominant compared to that of the derivative term (note that the magnitude of $P_{\epsilon_g\epsilon_m}$ can be also be altered by adjusting the scale under consideration). In line with the majority of existing literature \cite{Assassi:2015fma, Euclid:2024ris, Nikolis:2024kbx}, we choose to first focus on the terms proportional to $f_{\rm NL}$, as they are the most relevant for probing PNG. At the same time, we treat the $f_{\rm NL}P_{\epsilon_g\epsilon_m}$-order terms as the corrections to the leading-order PNG contribution and neglect all contributions of order $f_{\rm NL}^{2}$ and higher. As will be demonstrated in the following section, relative to the case of Gaussian initial conditions, the first-order saddle-point corrections to the conditional likelihood is proportional to $f_{\rm NL}P_{\epsilon_g\epsilon_m}$, whereas the corrections due to saddle-point fluctuations scales as $f_{\rm NL}$.


\section{Precision calculation of the EFT likelihood}
\label{sec:eftlikelihood}

Starting from this stage, we will formally proceed with the detailed calculation of the conditional likelihood under non-Gaussian initial conditions. As seen from eqs.~\eqref{pngmatter-partitionfunction} and \eqref{pngjoint-partitionfunction}, the inclusion of PNG introduces the same interaction term to both partition functions. In \cite{Cabass:2019lqx}, the authors considered the case where only the noise in the galaxy power spectrum $P_{\epsilon_g}$ is present and argued that, in this context, the non-Gaussian term contributes identically to the extra terms in both likelihoods. Consequently, when evaluating the ratio of the two, corresponding to the conditional likelihood, their contributions precisely cancel, yielding the same result as was previously obtained. Thus, in order to obtain nontrivial results, in this paper, we will focus on the case where both $P_{\epsilon_g}$ and $P_{\epsilon_g\epsilon_m}$ are present. We will illustrate through detailed calculations that the additional contributions to the two likelihoods originating from the saddle points do not cancel each other completely. Moreover, when accounting for the contributions around the saddle points, a field dependent, irreducible factor will also be introduced in the final conditional likelihood. 

\subsection{Set up the likelihood expressions}

We begin by focusing on the explicit formulae of the matter likelihood and joint likelihood. As discussed in \cite{Ke:2025uou}, following an appropriate reformulation, the expression for the likelihood can be rewritten as a path integral in three-dimensional Euclidean space. Firstly, the matter likelihood eq.~\eqref{pngmatter-likelihood} can be reformulated as follows
\begin{equation}
    \mathcal{P}[\delta] = \int\mathcal{D}\bphi \, e^{-S[\bphi]} \, , \label{likelihood1}
\end{equation}
where we have packaged all the integral variables into the field $\bphi$, such that
\begin{equation}
    \boldsymbol{\phi} = \begin{pmatrix}
        X \\ \deltain
    \end{pmatrix},
\end{equation}
In this manner, all terms in the exponential can be expressed in terms of the field $\bphi$, and subsequently interpreted as the action of the theory
\begin{align}
    S[\bphi]= \int_{\bk} \mathcal{J}^{a}(\bk) \bphi^{a}(-\bk) &+\frac{1}{2}\int_{\bk,\bk'}\mathcal{M}^{ab}(\bk,\bk') \bphi^{a}(\bk)\bphi^{b}(\bk') \nonumber\\ &+ \frac{1}{3!}  \int_{\bk,\p_1,\p_2} \mathcal{T}^{abc}(\bk,\p_{1},\p_{2}) \bphi^{a}(\bk)\bphi^{b}(\p_1)\bphi^{c}(\p_2) \, , \label{pngmatteraction}
\end{align}
where the cubic term in $\bphi$ arises as a consequence of the introduction of PNG. By comparing eq.~\eqref{pngmatteraction} with the terms in the partition function, we can straightforwardly derive the explicit expressions for the matrices
\begin{equation}
    \mathcal{J}^{a} = (-i\delta, 0) \; , \;  \mathcal{M}^{ab} = (2\pi)^{3} \delta_{D}(\bk+\bk')\begin{pmatrix}
        0 & iD_1 \\ iD_1 &P_{\rm in}^{-1} (k)
    \end{pmatrix} \; , 
    \end{equation}
    and
    \begin{equation}
    \mathcal{T}^{abc} = \left\{
\begin{array}{ll}
(2\pi)^3\delta_{D}(\bk + \p_{12})f_{NL}B(\bk,\p_1,\p_2)\;, & \text{for}\; a=b=c=2 \,, \\
0\;, & \text{otherwise }\,,
\end{array}
\right.
\end{equation}
where we have employed the approximate expansion of the forward matter field $\delta_{\rm fwd} [\deltain] \approx D_1 \deltain$ as taken in \cite{Cabass:2019lqx}. This reformulation implies that the calculation of the matter likelihood can be recast as the evaluation of a Euclidean path integral over multiple fields. Similarly, the expression for the joint likelihood eq.~\eqref{pngjoint-likelihood} can also be rewritten in an essentially identical formula
\begin{equation}
        \mathcal{P}[\delta_g,\delta] = \int\mathcal{D}\bphi_g \, e^{-S_g[\bphi_g]} \, . \label{likelihood2}
\end{equation}
In this context, we have preformed a similar packaging of the integral variables $\bphi_g = (X_g,X,\deltain)^{T}$. The action at this stage can be expressed as
\begin{align}
    S_g[\bphi_g] = \int_{\bk} \mathcal{J}_{g}^{a}(\bk) \bphi_{g}^{a}(-\bk) &+\frac{1}{2}\int_{\bk,\bk'}\mathcal{M}_{g}^{ab}(\bk,\bk') \bphi_{g}^{a}(\bk)\bphi_{g}^{b}(\bk') \nonumber\\ &+ \frac{1}{3!}  \int_{\bk,\p_1,\p_2} \mathcal{T}_{g}^{abc}(\bk,\p_{1},\p_{2}) \bphi_{g}^{a}(\bk)\bphi_{g}^{b}(\p_1)\bphi_{g}^{c}(\p_2) \, , \label{pngjointaction}
\end{align}
The expressions for the matrices can be derived in the same manner
\begin{equation}
    \mathcal{J}_{g} = (-i\delta_g,-i\delta,0) \;,\; \mathcal{M}_g = (2\pi)^3 \delta_{D}(\bk+\bk') \begin{pmatrix}
        P_{\epsilon_g}(k) & P_{\epsilon_g\epsilon_m}(k) & iD_1b_1 \\ P_{\epsilon_g\epsilon_m}(k) & 0 & iD_1 \\ iD_1b_1 &iD_1& P_{\rm in}^{-1}(k) 
    \end{pmatrix} \;.
\end{equation}
Similarly, we have only considered the leading-order bias expansion $\delta_{g, \rm fwd}[\deltain] \approx D_1 b_1 \deltain$. The non-Gaussian term now is
 \begin{equation}
    \mathcal{T}_{g}^{abc} = \left\{
\begin{array}{ll}
(2\pi)^3\delta_{D}(\bk + \p_{12})f_{NL}B(\bk,\p_1,\p_2)\;, & \text{for}\; a=b=c=3 \,,\\
0\;, & \text{otherwise }\,.
\end{array}
\right. 
\end{equation}
The primary objective of this paper is to compute the two likelihoods eq.~\eqref{likelihood1},  eq.~\eqref{likelihood2}, as well as the conditional likelihood defined by their ratio. Evidently, this requires us to employ the saddle-point expansion method discussed in section~\ref{sec:review} to perform the calculation. We will demonstrate that the implementation of PNG leads to nontrivial additional contributions to the conditional likelihood. In accordance with the previous spirit, we should decompose the path integral into contributions from the saddle points and their vicinity. We will now proceed to discuss these contributions individually.

\subsection{The contributions from the saddle points}

We now turn to the contributions from the saddle points to the full path integral. In fact, by mimicking the Gaussian-case calculations and inheriting their results, we can employ certain techniques to simplify the computations. We begin with the classical equations of motion of the two actions
\begin{align}
    \frac{\delta S[\bphi]}{\delta \bphi} \bigg|_{\bphi = \bar{\bphi}} = \mathcal{J} ^{a} + \mathcal{M}^{ab}\bar{\bphi}^{b} +\frac{1}{2} \mathcal{T}^{abc} \bar{\bphi}^{b} \bar{\bphi}^{c} = 0  \, , \label{saddle1}\\
    \frac{\delta S_g[\bphi_g]}{\delta \bphi_g} \bigg|_{\bphi_g = \bar{\bphi}_g} = \mathcal{J}_{g} ^{a} + \mathcal{M}_{g}^{ab}\bar{\bphi}_{g}^{b} +\frac{1}{2} \mathcal{T}_{g}^{abc} \bar{\bphi}_{g}^{b} \bar{\bphi}_{g}^{c} = 0  \, .  \label{saddle2}
\end{align}
It is evident that both of these saddle-point equations are quadratic matrix equations, and therefore their exact solutions typically involve the application of some eigenvalue analysis techniques. Fortunately, in our case, the matrices $\mathcal{T}$ and $\mathcal{T}_g$ can be treated as perturbations, as they are proportional to the PNG parameter $f_{\rm NL}$. Next, we will discuss separately the computation of the matter likelihood and the joint likelihood, along with their leading-order contributions corresponding to PNG.

\subsection*{The matter likelihood}

In the spirit of this approach, we first consider eq.~\eqref{saddle1} as an example, and decompose the saddle-point solution into the following form
\begin{equation}
    \bar{\bphi} = \bphi_0 + \bphi_1 \, , \label{phidecompose}
\end{equation}
where $\bphi_0$ represents the zeroth-order solution of eq.~\eqref{saddle1}, which can yield the same result presented in \cite{Cabass:2019lqx, Ke:2025uou}. This solution satisfies the equation
\begin{equation}
    \mathcal{J}^{a} + \mathcal{M}^{ab}\bphi_{0}^{b} = 0 \Rightarrow \bphi_{0} = \mathcal{M}^{-1}\mathcal{J} \equiv \begin{pmatrix}
        X_{\rm cl} \\ \delta_{\rm in,cl}
    \end{pmatrix}
    =\begin{pmatrix}
        \frac{i\delta}{D_{1}^{2}P_{\rm in}} \\ \frac{\delta}{D_1} 
    \end{pmatrix} \, , \label{phi0order}
\end{equation}
and $\bphi_1$ represents the first-order $f_{\rm NL}$ correction to the solution. Then the current situation is completely analogous to what is encountered in perturbation theory within quantum mechanics. Substituting the expansion of the solution eq.~\eqref{phidecompose} into eq.~\eqref{saddle1}, and making use of the zeroth-order equation eq.~\eqref{phi0order}, we obtain
\begin{equation}
    \mathcal{M}^{ab} \bphi_{1}^{b} + \frac{1}{2}\mathcal{T}^{abc} \bphi_{0}^{b} \,\bphi_{0}^{c} =0\,.
\end{equation}
Thus, we can obtain the expression for the perturbed field
\begin{equation}
    \bphi_{1}^{a} = -\frac{1}{2}(\mathcal{M}^{-1})^{ab} \,\mathcal{T}^{bcd} \bphi_{0}^{c}\, \bphi_{0}^{d} = \begin{pmatrix}
        \frac{iBf_{\rm NL}\delta^2}{2D_{1}^{3}} \\ 0
    \end{pmatrix} \, . \label{phi1order}
\end{equation}
By iteratively employing this method, we can obtain solutions to arbitrary higher orders. It is straightforward to observe that, after incorporating the leading-order PNG, the corrected saddle-point solution compared to eq.~\eqref{phi0order} effectively corresponds to a shift of the matter current $X\rightarrow X + \frac{iBf_{\rm NL}\delta^2}{D_{1}^{3}} $, and this correction is hierarchically suppressed by powers of $f_{\rm NL}$. This property enables us to carry over the discussion and results on the saddle-point contribution from \cite{Cabass:2019lqx}, and to explore additional terms induced by the shift of $X$. Specifically, the terms in the action that give rise to corrections at order $f_{\rm NL}$ due to this adjustment are
\begin{equation}
    S[\bar{\bphi}] \supset \exp\left\{ \int_{\bk}\left(-iX\delta -iD_1X\deltain\right) +\int_{\bk,\p_1,\p_2}f_{NL}B(\deltain)^3\right\} \bigg|_{\bphi=\bar{\bphi}}\,,
\end{equation}
where we have omitted the momentum dependence of each term and the Dirac delta function in this expression for convenience. Plugging eqs.~\eqref{phi0order} and \eqref{phi1order} into eq.~\eqref{phidecompose}, we can read off from the action that
\begin{align}
    -i \int_{\bk} X(\bk) \delta(-\bk) &\supset \frac{f_{\rm NL}}{2D_{1}^{3}} \int_{\bk,\p_1,\p_2} \delta(-\bk) B(\bk,\p_1,\p_2) \delta(\p_1)\delta(\p_2) + \mathcal{O}(f_{\rm NL}^{2}) \, ,\\
   - iD_{1}\int_{\bk} X(\bk)\deltain(-\bk) &\supset  \frac{f_{\rm NL}}{2D_{1}^{3}} \int_{\bk,\p_1,\p_2} \delta(-\bk) B(\bk,\p_1,\p_2) \delta(\p_1)\delta(\p_2) + \mathcal{O}(f_{\rm NL}^{2}) \,, \\ 
    \frac{1}{3!}\int_{\bk,\p_1,\p_2} \delta_{D}(\bk + \p_{12})&B(\bk,\p_1,\p_2) \deltain(\bk)\deltain(\p_1)\deltain(\p_2)\supset \nonumber\\ &\frac{f_{\rm NL}}{3!D_{1}^{3}} \int_{\bk,\p_1,\p_2} \delta(\bk) B(\bk,\p_1,\p_2)\delta(\p_1)\delta(\p_2)+ \mathcal{O}(f_{\rm NL}^{2}) \,.
\end{align}
The aforementioned terms correspond to the supplementary contributions to the matter likelihood upon the introduction of PNG. Note that in calculating these contributions, we must consider not only the terms in the action that depend on $X$, but also the non-Gaussian effective interaction. By employing the same method, we can also extend this calculation to arbitrary higher orders.

\subsection*{The joint likelihood}

We now examine a more complicated scenario concerning the joint likelihood. Given the complexity of its formulation, we will first consider the simplest case, where $P_{\epsilon_g \epsilon_m} = 0$, and subsequently restore the general case. By assuming the same field-decomposition ansatz and following an analogous derivation process, we can show that
\begin{equation}
    \bar{\bphi}_g = \bphi_{g,0} + \bphi_{g,1} \, , \; \bphi_{g,1} = -\frac{1}{2} (\mathcal{M}_{g}^{-1})^{ab} \mathcal{T}_{g}^{bcd} \bphi_{0,g}^{c} \bphi_{0,g}^{d} = \begin{pmatrix}
        0 \\ \frac{iBf_{\rm NL}\delta^2}{2D_{1}^{3}}\\0
    \end{pmatrix}\, . \label{jointsaddle}
\end{equation}
From this expression, it is evident that in this context, to obtain the additional terms induced by PNG, we can still perform a shift $X_{\rm cl}\rightarrow X_{\rm cl} + \frac{iBf_{\rm NL}\delta^2}{D_{1}^{3}}$, substitute it into the action, and then isolate the terms proportional to $f_{\rm NL}$. At the same time, if we recover $P_{\epsilon_g\epsilon_m}$, the perturbative solution $\bphi_{g,1}$ should be modified to 
\begin{equation}
    \bphi_{g,1}' = \frac{Bf_{\rm NL}}{2P_{\epsilon_g}D_{1}^{2} } \begin{pmatrix}
        -iD_1P_{\epsilon_g\epsilon_m} \left(\frac{\delta}{D_1} + \frac{iP_{\epsilon_g\epsilon_m}X_{g, \rm cl}}{D_1}\right)^2 \\ 
        (iD_1P_{\epsilon_g}-ib_1D_1P_{\epsilon_g\epsilon_m})\left(\frac{\delta}{D_1} + \frac{iP_{\epsilon_g\epsilon_m}X_{g, \rm cl}}{D_1}\right)^2 \\ P_{\epsilon_g\epsilon_m}^{2}\left(\frac{\delta}{D_1} + \frac{iP_{\epsilon_g\epsilon_m}X_{g, \rm cl}}{D_1}\right)^2 
    \end{pmatrix}\, , \label{jointsaddle1}
\end{equation}
where $X_{g,\rm cl}$ is defined via
\begin{equation}
    X_{g, \rm cl} = \frac{i(\delta_g-b_1\delta)}{P_{\epsilon_g}}\, . 
    \label{xg}
\end{equation}
When taking $P_{\epsilon_g\epsilon_m} = 0$ we will recover the result of eq.~\eqref{jointsaddle}. 

By comparing eqs.~\eqref{jointsaddle} and \eqref{jointsaddle1}, we can observe that after recovering $P_{\epsilon_g\epsilon_m}$, in addition to reproducing the same shift as before, all remaining terms are at least proportional to $P_{\epsilon_g\epsilon_m}$ or $f_{\rm NL}$. On the other hand, from our previous discussion, we know that $P_{\epsilon_g\epsilon_m} \sim k^2$, implying that its contribution is suppressed at large scales. This enables us to treat it as a perturbative parameter similar to $f_{\rm NL}$. Note that the leading-order saddle-point solution \cite{Ke:2025uou},
\begin{equation}
      \bar{\bphi}_{g,0} = \begin{pmatrix}
        X_{g,\rm cl} + \frac{P_{\epsilon_g \epsilon_m}}{P_{\epsilon_g}}(2b_{1}X_{g,\rm cl} - \frac{i\delta}{D_{1}^{2}P_{\rm in}}) \\
        \frac{i\delta}{D_{1}^{2}P_{\rm in}} - b_1X_{g, \rm cl} -\frac{b_1P_{\epsilon_g \epsilon_m}}{P_{\epsilon_g}}(2b_{1}X_{g, \rm cl} - \frac{i\delta}{D_{1}^{2}P_{\rm in}}) -\frac{P_{\epsilon_g \epsilon_m}X_{g, \rm cl}}{D_{1}^{2}P_{\rm in}} \\
        \frac{\delta}{D_1} + \frac{iP_{\epsilon_g \epsilon_m}X_{g, \rm cl}}{D_1} 
    \end{pmatrix}\,, \label{bar-phig}
\end{equation}
inherently involves $P_{\epsilon_g\epsilon_m}$. Therefore, if we include the $P_{\epsilon_g\epsilon_m}$-related components in $\bphi_{g,1}$, many of the resulting terms naturally correspond to higher-order contributions. In the present work, our objective is to isolate the leading-order contributions proportional to $P_{\epsilon_g\epsilon_m}$ and $f_{\rm NL}$. Accordingly, it is reasonable to approximate the first-order perturbed solution as follows
\begin{equation}
     \bphi_{g,1}'' =  \frac{Bf_{\rm NL}}{2D_1^{3}}\begin{pmatrix}
         -\frac{iP_{\epsilon_g\epsilon_m}\delta^2}{P_{\epsilon_g}} \\ i\delta^2 -2P_{\epsilon_g\epsilon_m}\delta X_{g,\rm cl} - \frac{ib_1P_{\epsilon_g\epsilon_m}\delta^2}{P_{\epsilon_g}} \\ 0
     \end{pmatrix} \,.
\end{equation}
This expression implies that both the galaxy current and the matter current in the action should be shifted accordingly
\begin{align}
    X_{g,} &\rightarrow X_{g} - \frac{iBf_{\rm NL}P_{\epsilon_g\epsilon_m}\delta^2}{2P_{\epsilon_g}D_{1}^{3}} \,, \\  \,X &\rightarrow X  + \frac{Bf_{\rm NL}}{2D_{1}^{3}} \left(i\delta^2 -2P_{\epsilon_g\epsilon_m}\delta X_{g,\rm cl} - \frac{ib_1P_{\epsilon_g\epsilon_m}\delta^2}{P_{\epsilon_g}}\right)  \, ,
\end{align}
and the expression for the classical initial matter field remains unchanged. Following this adjustment, the terms in the action that can potentially generate contributions proportional to $P_{\epsilon_g\epsilon_m}$, $f_{\rm NL}$ and $P_{\epsilon_g\epsilon_m}f_{\rm NL}$ are given by
\begin{align}
    S_g[\bphi_g] \supset &\exp\left\{\int_{\bk}(-iX\delta-iX_g\delta_g - iD_1X\deltain - ib_1D_1X_g\deltain) \right\} \nonumber \\&\times \exp \left\{- \int_{\bk} P_{\epsilon_g\epsilon_m}X_gX+ \frac{f_{\rm NL}}{3!}\int_{\bk,\p_1,\p_2} B (\deltain)^3\right\} \bigg|_{\bphi_g = \bar{\bphi}_g}\,,
\end{align}
In order to prevent any potential omission, all terms in the action should be treated carefully throughout the computation. Although the resulting expression is complicated, we present its full form below: 

First, there are terms arising from the functional Fourier transform
\begin{align}
    -i \int_{\bk} &X \delta \supset \frac{f_{\rm NL}}{2D_{1}^{3}} \int_{\bk,\p_1,\p_2} B(\bk,\p_1,\p_2)\left(\delta(\p_1)\delta(\p_2)+2iP_{\epsilon_g\epsilon_m}(p_1)\delta(\p_1)X_{g,\rm cl}(\p_2) \right) \delta(-\bk)
  \nonumber\\& \; \qquad \qquad 
    -\frac{b_1f_{\rm NL}}{2D_{1}^{3}} \int_{\bk,\p_1,\p_2} B(\bk,\p_1,\p_2)\left(\frac{P_{\epsilon_g\epsilon_m}(k)\delta(\p_1)\delta(\p_2)}{P_{\epsilon_g}(k)}\right)\delta(-\bk) \, ,\\
    -i\int_{\bk} &X_g\delta_g \supset - \frac{f_{\rm NL}}{2D_{1}^{2}}\int_{\bk,\p_1,\p_2} B(\bk,\p_1,\p_2) \frac{P_{\epsilon_g\epsilon_m}(k)}{P_{\epsilon_g}(k)} \delta(\p_1)\delta(\p_2) \delta_{g}(-\bk) \, .
    \end{align}
   Next, there are other terms proportional to the currents in the action
    \begin{align}
    -iD_1 \int_{\bk} &X\deltain\supset \frac{f_{\rm NL}}{2D_{1}^{3}}\int_{\bk,\p_1,\p_2} B(\bk,\p_1,\p_2)\delta(\p_1) \delta(\p_2)\left(\delta +i P_{\epsilon_g\epsilon_m}X_{g, \rm cl}\right)(-\bk) \nonumber \\
    +\frac{f_{\rm NL}}{2D_{1}^{3}}&\int_{\bk,\p_1,\p_2} B(\bk,\p_1,\p_2) \left(2iP_{\epsilon_g\epsilon_m}(p_1)\delta(\p_1)X_{g,\rm cl}(\p_2) -\frac{P_{\epsilon_g\epsilon_m}(k)}{P_{\epsilon_g}(k)}\delta(\p_1)\delta(\p_2)\right)\delta(-\bk) \, , \\ -ib_1D_1\int_{\bk} & X_g \deltain \supset -\frac{b_1f_{\rm NL}}{2D_{1}^{3}} \int_{\bk,\p_1,\p_2} B(\bk,\p_1,\p_2) \frac{P_{\epsilon_g\epsilon_m}(k)}{P_{\epsilon_g}(k)} \delta(\p_1)\delta(\p_2)\delta(-\bk) \, .
    \end{align}
    Then, there is a term related to the noise terms
    \begin{align}
    -\int_{\bk}&P_{\epsilon_g\epsilon_m}X_{g}X \supset -\frac{-if_{\rm NL}}{2D_{1}^{3}} \int_{\bk,\p_1,\p_1} B(-\bk,\p_1,\p_2)P_{\epsilon_g\epsilon_m}(k) X_{g,\rm cl}(k) \delta(\p_1)\delta(\p_2) \, .
    \end{align}
    Finally, there is a non-Gaussian interaction term
    \begin{align}
    \frac{1}{3!}&\int_{\bk,\p_1,\p_2}B(\bk,\p_1,\p_2)\deltain(\bk)\deltain(\p_1)\deltain(\p_2) \supset \nonumber 
    \frac{1}{2D_{1}^{3}}\int_{\bk,\p_1,\p_2} \delta_{D}(\bk + \p_{12})\nonumber \\ & B(\bk,\p_1,\p_2) (\delta + iP_{\epsilon_g\epsilon_m}X_{g, \rm cl})(\bk)(\delta + iP_{\epsilon_g\epsilon_m}X_{g, \rm cl})(\p_1)(\delta + iP_{\epsilon_g\epsilon_m}X_{g, \rm cl}) (\p_2) \, .
\end{align}
A comparison of the matter and joint likelihood expressions shows that all irreducible saddle-point contributions are proportional to $f_{\rm NL}P_{\epsilon_g\epsilon_m}$, in agreement with the discussion in \cite{Cabass:2019lqx}. Because $P_{\epsilon_g\epsilon_m} \propto k^2$ corresponds to higher-derivative (gradient) terms and $f_{\rm NL}$ itself is small, these new terms are subject to a double suppression. Based on our previous estimates of the magnitudes of $f_{\rm NL}$ and $P_{\epsilon_g\epsilon_m}$, we find that the $P_{\epsilon_g\epsilon_m}$ contribution is typically subdominant to that of $f_{\rm NL}$. Consequently, the terms derived here are smaller than the $P_{\epsilon_g\epsilon_m}$ terms retained in \cite{Cabass:2019lqx} but larger than the neglected $P_{\epsilon_g\epsilon_m}^{2}$ contributions. Moreover, recent Bayesian forward-modeling analyses at the field level already include $f_{\rm NL}k^2$ terms -- for example, the galaxy-bias expansion in \cite{Euclid:2024ris} is carried out to this order. We therefore present these terms explicitly, both for completeness and as a pedagogical reference for future applications. At the same time, one might be concerned that certain terms in the action contain explicit factors of the imaginary unit $i$, potentially rendering the expression complex. However, this is not the case: due to the presence of $i$ in the definition of $X_{g, \rm cl}$ itself in eq.~\eqref{xg}, all imaginary components in $S_g[\bar{\phi}_g]$ cancel out exactly, ensuring that the resulting expression is manifestly real -- consistent with the physical interpretation of probability. 

A brief comparison of the expressions for the two likelihoods with those under Gaussian initial conditions reveals certain underlying structures. We can observe that when $P_{\epsilon_g\epsilon_m} = 0$, the contributions of PNG to both likelihoods are exactly the same, consistent with the discussion presented in \cite{Cabass:2019lqx}. Conversely, when the cross stochastic term is introduced, its contributions to the joint likelihood prevent this cancellation when computing the ratio of the two likelihoods, thereby imparting non-vanishing contributions to the conditional likelihood. The relationship between these noise terms and the likelihood expressions is of considerable interest, and we will discuss it in detail in the following subsection.

\subsection{The contributions around the saddle points}

We now turn to the section on precision calculations -- specifically, employing the saddle-point expansion method to compute the contributions around the saddle points. To achieve this, we decompose the fields $\bphi$ and $\bphi_g$ into their saddle-point solutions and some small fluctuations around them
\begin{equation}
   \bphi = \bar{\bphi} + \bvarphi \, , \;\qquad \bphi_g = \bar{\bphi}_{g} + \bvarphi_g \, .  \label{fieldexpansion}
\end{equation}
It is important to note that we should not confuse the expansion of the field around the saddle points eq.~\eqref{fieldexpansion} with the perturbative expansion of the saddle-point solutions eqs.~\eqref{phidecompose} and \eqref{jointsaddle}. The former refers to decomposing the integral into contributions from the saddle points and their neighborhoods, while the latter refers to the contributions at the saddle points, ordered from lower to higher orders. We will continue to consider the two cases separately. First, the field decomposition leads to the following expression for the matter likelihood
\begin{equation}
    \mathcal{P}[\delta] = \int\mathcal{D}\bphi \, e^{-S[\bphi]} \approx  e^{-S[\bar{\bphi}]} \int \mathcal{D}\bvarphi\, e^{-\frac{1}{2}S''[\bar{\bphi}]\bvarphi^2} \, .
\end{equation}
To compute this integral, we need to determine the eigenvalues of the second derivative of the action at the saddle point. By substituting the saddle-point solution eq.~\eqref{phidecompose}, we can obtain
\begin{equation}
    S''[\bar{\bphi}] = \mathcal{M}^{ab} + \mathcal{T}^{abc} \bar{\bphi}^{c} 
 = \begin{pmatrix}
     0  & iD_1 \\ iD_1 & P_{\rm in}^{-1} + \frac{f_{\rm NL}B\delta}{D_1}
 \end{pmatrix} \, , 
\end{equation}
Since the matrix $\mathcal{T}$ is proportional to $f_{\rm NL}$, a simplification arises whereby it suffices to substitute only the leading-order saddle-point solution $\bphi_0$ into $\bar{\bphi}$. We now need to determine the eigenvalues of this matrix. By comparing its structure with the formula presented in \cite{Ke:2025uou}, we observe that, after an appropriate redefinition of the variable
\begin{equation}
    P_{\rm in}'^{-1} \equiv P_{\rm in}^{-1} + \frac{f_{\rm NL}B\delta}{D_1} \, ,
\end{equation}
the expression for $S''[\bar{\bphi}]$ becomes formally identical to that derived in \cite{Ke:2025uou}. In this way, we can directly invoke the previously established result, whereby the corresponding eigenvalues are given by
\begin{equation}
    m_1 = \frac{1}{2}(P_{\rm in}'^{-1} - \sqrt{-4 D_{1}^{2} +P_{\rm in}'^{-2}}) \, , \,\,\,\,\,\, m_2 =  \frac{1}{2}(P_{\rm in}'^{-1} + \sqrt{-4 D_{1}^{2} +P_{\rm in}'^{-2}})\,.
\end{equation}
It is evident that if the two eigenvalues are real, they must also be positive, as the growth factor $D_1$ can be consistently expected to be positive. Therefore, there are no concerns regarding negative-mode solutions at this stage. We can thus safely proceed with the saddle-point expansion method and apply the Gaussian integration formula, leading to the following result:
\begin{equation}
      \mathcal{P}[\delta] \approx   e^{-S[\bar{\bphi}]}({\rm det}S''[\bar{\bphi}])^{-1/2} =\sqrt{\pi}\,(\prod_i m_i)^{-1/2} \, {\rm e}^{-S[\bar{\bphi}]}=\frac{\sqrt{\pi}}{D_1}\,  e^{-S[\bar{\bphi}]}\,. 
    \label{matter-answer}
\end{equation}
Here, when plugging in $S[\bar{\bphi}]$, we have already incorporated the additional contributions arising from the inclusion of PNG. In this case, it is evident that although the eigenvalues $m_n$ exhibit explicit dependence on the matter field $\delta$, such dependence cancels out in their product. Consequently, the resulting matter likelihood precisely matches that obtained in \cite{Ke:2025uou}. As established in the previous discussions, this field-independent prefactor will be consistently absorbed into the normalization factor associated with the definition of functional Dirac delta function, and therefore it does not yield any nontrivial contribution to the matter likelihood.

We now turn our attention to the joint likelihood, whose formulation is structurally analogous to that of the matter likelihood
\begin{equation}
     \mathcal{P}[\delta_g,\delta] = \int \mathcal{D} \bphi_g \, {\rm}e^{-S_g[\bphi_g]}
\approx \int \mathcal{D} \bvarphi_g \, e^{-S_g[\bar{\bphi}_g] - \frac{1}{2}S_g''[\bar{\bphi}_g]\bvarphi_g^2 }\,.
\end{equation}
Similarly, the matrix whose eigenvalues are to be determined in this case is given by
\begin{equation}
    S_g''[\bar{\bphi}_g] = \mathcal{M}_{g}^{ab} +\mathcal{T}_{g}^{abc}\bar{\bphi}_{g}^{c} = \begin{pmatrix}
        P_{\epsilon_g} & P_{\epsilon_g\epsilon_m} & iD_1b_1 \\ P_{\epsilon_g\epsilon_m} &0 &iD_1
 \\ iD_1b_1  &iD_1 &P_{\rm in}''^{-1}    \end{pmatrix}\,,
\end{equation}
where we have already defined
\begin{equation}
    P_{\rm in}''^{-1}  = P_{\rm in}^{-1} + f_{\rm NL}B\left(\frac{\delta}{D_1} - \frac{iP_{\epsilon_g\epsilon_m}X_{g, \rm cl}}{D_1}\right)\,. \label{jointshift}
\end{equation}
Even though we have employed such a redefinition to simplify our calculation, $S_g''[\bar{\bphi}_g]$ remains a three-dimensional matrix and may still possess negative or complex eigenvalues, thereby rendering the corresponding functional integral ill-defined. Accordingly, when computing the joint likelihood, it is essential to follow the strategy employed in section~\ref{sec:review}, complexifying the action and identifying the combination of the steepest descent contours in the complex plane that is homologous to the original integration contour (the real line in the field space). The full path integral can then be approximated by summing over all the contributions from these complex contours, which ensures a well-defined and convergent result. All relevant technical details have been provided in the preceding sections, and here we directly employ the gradient flow equation to determine the corresponding steepest descent contours. Assuming that the same steps as previously outlined have already been performed, the gradient flow equation in this case can be written as
\begin{equation}
   \frac{\partial \bphi_g(\bk,u)}{\partial u} =  \mathcal{J}_{g}^{*a} + \mathcal{M}_{g}^{*ab}\bphi_{g}^{*b} + \frac{1}{2}\mathcal{T}_{g}^{*abc}\bphi_{g}^{*n}\bphi_{g}^{*c} \,.  
\end{equation}
To solve this equation, we employ the saddle-point decomposition of the field eq.~\eqref{fieldexpansion} and the complex conjecture of the saddle-point equation eq.~\eqref{saddle2}. Under this procedure, the equation can be recast as follows
\begin{equation}
    \frac{\partial\bvarphi_g(\bk,u)}{\partial u} = (\mathcal{M}_{g}^{*ab} + \mathcal{T}_{g}^{*abc}\bar{\bphi}_{g}^{*c}) \bvarphi_{g}^{*b}\, ,
\end{equation}
where we have used the fact that $\bar{\bphi}_g$ is independent of $u$, since $u$ simply serves as a parameter labeling different integration contours through the saddle point in the complex plane. At this time, the gradient flow equation is modified from its previous form eq.~\eqref{egradient} by the addition of a non-Gaussian term, which is precisely equivalent to implementing the variable shift eq.~\eqref{jointshift} (note that $P_{\rm in}''^{-1}$ is real). Therefore, by implementing this shift, one can construct the same eigen-equation and convert it into a Hermitian eigen-equation as in section \ref{sec:review}, with the expressions remaining identical to those derived previously. Consequently, the sum over these positive and real eigenvalues $m_n$ can be obtained
\begin{equation}
      \sum_{n} m_n =m_1+m_2+m_3 = |P_{\epsilon_g} - P_{\rm in}''^{-1}| = \bigg|P_{\epsilon_g} - P_{\rm in}^{-1} - f_{\rm NL}B\left(\frac{\delta}{D_1} - \frac{iP_{\epsilon_g\epsilon_m}X_{g, \rm cl}}{D_1}\right)\bigg| \,. \label{sum-of-pngmn}
\end{equation}
From this point forward, we will use the simplified notation $P_{\rm in}''^{-1}$, with the understanding that it depends on the fields $\delta$ and $\delta_g$.  With these elements established, we are now ready to compute the corresponding joint likelihood
\begin{align}
    \mathcal{P}_g[\delta_g,\delta] & \approx  \int_{C} \mathcal{D} \boldsymbol{\phi}_g \,  e^{-S_g[\bar{\bphi}_g] -\frac{1}{2} \bvarphi_g S''_g[\bar{\bphi}_g]\bvarphi_g}  = \int_C \mathcal{D} \bvarphi_g \,  e^{-S_g[\bar{\bphi}_g] - \bvarphi_g (\mathcal{M}_g + \mathcal{T}_g \bar{\bphi}_g)\bvarphi_g}\nonumber\\ 
    &   =e^{-S_g[\bar{\bphi}_g]} \frac{\pi}{\sum_n m_n} \approx e^{-S_g[\bar{\bphi}_g]} \frac{\pi}{|P_{\epsilon_g}-P_{\rm in}''^{-1}|}.   \label{pngjoint-result}
\end{align}
We can see that the resulting expression contains not only the saddle-point contributions but also an additional field-dependent prefactor. Furthermore, fluctuations around the saddle point generate a non-trivial contribution to the joint likelihood that is proportional to to $f_{\rm NL}$. This term is the leading-order correction sourced exclusively by PNG, and is therefore of particular importance. We will discuss it in detail in a later section. At this stage, the prefactor exhibits an explicit dependence on both the matter and galaxy fields, and cannot be eliminated by setting $P_{\epsilon_g\epsilon_m} = 0$. In other words, once the contributions of PNG are taken into account, the precision calculation process introduces a field-dependent term that cannot be absorbed into the normalization procedure. Consequently, this term appears explicitly in the final expression for the conditional likelihood. Its potential implications will be examined in detail in the following subsection.

\subsection{The conditional likelihood}

We are now exceedingly close to the final result, yet a few remaining preparatory steps must still be completed. We denote the logarithms of the two likelihoods as $\wp[\delta]$ and $ \wp[\delta_g,\delta]$, and discard all field-independent terms. This yields
\begin{align}
    \wp[\delta] &= {\rm ln} \,\mathcal{P}[\delta] = -S[\bar {\bphi}] \, , \label{matterlog} \end{align}
    \begin{align}
      \wp[\delta_g,\delta]&= {\rm ln}\,\mathcal{P}[\delta_g,\delta]  \approx-{\rm ln}\left(\frac{1}{\bigg|P_{\epsilon_g}-P_{\rm in}''^{-1}\bigg|}\right) S_g[\bar{\bphi}_g]\,.\label{jointlog}
\end{align}
Accordingly, the expression for the logarithmic conditional likelihood can be obtained by taking the ratio of the two likelihood functions
\begin{equation}
        \wp[\delta_g |\delta] = {\rm ln}\,\mathcal{P}[\delta_g|\delta] ={\rm ln}\, \frac{\mathcal{P}[\delta_g,\delta]}{\mathcal{P}[\delta]} = -S[\bar{\boldsymbol{\phi}}] +{\rm ln}\left(\frac{1}{\bigg|P_{\epsilon_g}-P_{\rm in}''^{-1}\bigg|}\right)S_g[\bar{\boldsymbol{\phi}}_g] \, .
    \label{conditional-loglikelihood}
\end{equation}
We now make some comments on the calculation procedure, the resulting expressions, and the potential implications and insights that may be inferred from them:
\begin{itemize}
    \item We begin by examining these stochastic noise terms. It is evident that when $P_{\epsilon_g\epsilon_m} = 0$, the saddle-point contributions to both the matter likelihood and the joint likelihood exactly coincide, implying that the conditional likelihood upon introducing PNG remains the same as in the previous one. In most forward modeling approaches, considering only the saddle-point contributions typically provides a reasonable approximation, as noted in \cite{Cabass:2019lqx}. However, for more accurate theoretical predictions, it becomes essential to incorporate higher-order noise contributions into the analysis \cite{Cabass:2020nwf}. These noise terms are of critical importance, as they are introduced to cancel the UV dependence of the loop diagrams \cite{Cabass:2019lqx}. Consequently, their inclusion in the computation bears a natural analogy to the construction of the renormalized action in quantum field theory (and these noise terms can be interpreted as counterterms) \cite{Schwartz:2014sze}. Similarly, we can generalize the calculation to encompass higher-order scenarios, analogous to the 1PI \cite{Andreassen:2016cvx, Weinberg:1987vp} and 2PI \cite{Garbrecht:2015oea, Garbrecht:2015yza} effective action in quantum field theory. Furthermore, it is the cross stochasticity between the matter and galaxy fields that induces the inconsistency and lack of cancellation of their respective saddle-point contributions, resulting in nontrivial contributions to the conditional likelihood when PNG are incorporated. This indicates that even at the same perturbative order, we cannot directly cancel the identical perturbative terms between the two likelihoods during the computation. We conclude that considering higher-order noise terms introduces new contributions to the conditional likelihood.
    
    \item We next compare the contributions from the saddle points and around the saddle points. In our original expectation, the effective interaction term in the partition functions introduced by PNG is proportional to $\deltain^{3}$, as shown in eqs.~\eqref{pngmatter-partitionfunction} and \eqref{pngjoint-partitionfunction}, thereby yielding additional terms in the equations of motion eqs.~\eqref{saddle1}, \eqref{saddle2}. Thus, this contribution will be reflected in the field dependence of the saddle-point solution, which in turn induces a corresponding dependence of the fields on the eigenvalues of $S''[\bar{\bphi}]$ and $S_{g}''[\bar{\bphi}_g]$,  resulting in an irreducible factor in both likelihoods. Interestingly, when computing the matter likelihood, although all the eigenvalues exhibit explicit field dependence, this dependence cancels out upon performing the Gaussian integration, leading to trivial additional contributions. We observe that even when ignoring the cross stochastic term $P_{\epsilon_g\epsilon_m}$, the contributions around the saddle points differ fundamentally between the two likelihoods: one exhibits field dependence, whereas the other does not. Thus, we argue that the precision calculation in the presence of PNG will also yield nontrivial results even when only considering the lowest-order noise.
    
    \item Here, we describe how to incorporate more cosmological effects into our calculation. Clearly, these effects can be manifested in the corresponding terms within the partition functions. For instance, following the spirit of effective field theory, the implementation of PNG also necessitates adjustments to the galaxy bias expansion via the introduction of new effective expansion operators \cite{Assassi:2015fma}. These adjustments will be reflected in the reformulation of the expression for $\delta_{g,\rm fwd}[\deltain]$. We emphasize that in our calculations, although a simplified treatment of $\delta_{g, \rm fwd}$ is applied, none of the computational steps depend on its specific form. Therefore, when we want to incorporate this effect, we can directly modify $\delta_{g,\rm fwd}[\deltain]$ via the desired bias expansion and proceed with the calculation. Alternatively, if we seek to incorporate higher-order PNG, we would need to expand the non-Gaussian primordial potential, as given in eq.~\eqref{primordialpotential}, to higher orders. This generalization will only lead to higher-order vertices of $\deltain$ in the partition functions eqs.~\eqref{pngmatter-partitionfunction} and \eqref{pngjoint-partitionfunction}. Thus, we can still extract the additional terms associated with the saddle-point contributions by appropriately shifting the saddle-point solutions. Meanwhile, around the saddle points, these vertices will also result in modifications to $P_{\rm in}^{-1}$. Similarly, we can also incorporate higher-order galaxy stochastics, which correspond to higher-order terms of $J$ and $J_g$, and viscous effects, which correspond to $JJ\deltain$ term \cite{Cabass:2019lqx}, among others. The detailed corrections to the conditional likelihood induced by these effects will be left for future work to investigate.
\end{itemize}    

\subsection{Connections to Bayesian forward modeling}

We conclude this section with a discussion of the implication of our results for Bayesian forward modeling \cite{Jasche:2012kq, Wang:2014hia, Jasche:2018oym, Kitaura:2019ber, Bos:2018rpw}. In fact, Bayesian methods rely fundamentally on the conditional probability density functional $\mathcal{P}[\delta_g|\deltain]$ that describes the likelihood of observing the tracer field $\delta_g$, conditioned on a given realization of the initial conditions $\deltain$. According to \cite{Schmidt:2018bkr, Cabass:2019lqx}, we can generally decompose $\mathcal{P}[\delta_g|\deltain]$ into two quantities: i) the conditional likelihood $\mathcal{P}[\delta|\deltain]$, i.e. the probability of having a total (dark+luminous) matter field $\delta$ given the initial conditions $\deltain$. ii) the conditional likelihood $\mathcal{P}[\delta_g|\delta]$ of observing $\delta_g$ given a matter field $\delta$. The former is generally informed by forward models of the matter distribution (such as $N$-body simulations), wheres the latter often requires support form semi-analytical calculations. The aim of this work, along with that of several recent studies \cite{Schmidt:2018bkr, Cabass:2019lqx, Ke:2025uou, Voivodic:2025quw}, is to provide an analytical and precision computation of the latter.

In this work, we present the formal treatment of the impact of PNG on the conditional likelihood $\mathcal{P}[\delta_g|\delta]$, yielding two kinds of irreducible contributions in the final result: one arising from the saddle-point contributions, scaling as $k^2f_{\rm NL}$, and the other from fluctuations around the saddle points, which scale as $f_{\rm NL}$ and $k^2f_{\rm NL}$, respectively. However, it is important to note that a field-level analysis on PNG data necessarily requires incorporating non-Gaussian $\mathcal{P}[\delta|\deltain]$ in the first step, prior to applying the likelihood derived in this work \cite{Andrews:2022nvv, Euclid:2024ris}. Consequently, the effects of PNG are already partially embedded at this stage. As a result, even if a likelihood without PNG is employed in the second step, the inferred value of $f_{\rm NL}$ may be biased, yet it will not trivially converge to zero. To estimate the potential impact of our results on parameter interference, we expand eq.~\eqref{conditional-loglikelihood} perturbatively with respect to $f_{\rm NL}$,
\begin{align}
    \wp[\delta_g |\delta] &\approx -S[\bar{\bphi}] + S_g[\bar{\bphi}_g] - f_{\rm NL}B\left(\frac{\delta}{D_1} - \frac{iP_{\epsilon_g\epsilon_m}X_{g,\rm cl}}{D_1}\right) S_g[\bar{\bphi}_g]\nonumber \\ 
    & \sim \text{Gaussian results} \, + \underset{\rm saddle-point \, fluctuations}{\mathcal{O}(f_{\rm NL})} + \underset{\rm saddle \, point + fluctuations}{\mathcal{O}(k^2f_{\rm NL})}  + \cdots \, ,
    \end{align}
where `$\cdots$' denotes higher-order terms. Our analysis reveals that the leading contribution proportional to $f_{\rm NL}$, which arises from saddle-point fluctuations, dominates the corrections, and constitutes a unique prediction of the application of the saddle-point expansion method. On the other hand, the next-to-leading term, proportional to $k^2f_{\rm NL}$, is subject to a double suppression due to both its higher-derivative nature and the small amplitude of PNG. Nevertheless, recent PNG Bayesian forward modeling have begun to incorporate this order of contribution \footnote{This can be seen from the galaxy bias expansion present in \cite{Euclid:2024ris}.} to help us provide more precise constraints on PNG parameters. Accordingly, we choose to make the explicit expression of this order term manifest, in preparation for its potential relevance in future analyses. Compared to extracting primordial non-Gaussian signals from CMB anisotropies, probing such signatures directly from LSS is significantly more challenging due to non-linear fluctuations \cite{Baumann:2009ds}. We hope our work contribute to the development of more comprehensive forward models, enhance the accuracy of parameter constraints, and increase the flexibility of the future simulations.

\section{Summary and conclusion}
\label{sec:conclusion}

The conditional likelihood is the key physical quantity required when implementing Bayesian forward modeling. In this paper, we provide a functional formulation for the precision calculation of the EFT conditional likelihood, considering non-Gaussian initial conditions. The precision here is reflected in two aspects: first, we have considered the contribution of higher-order noise terms in the partition functions eqs.~\eqref{pngmatter-partitionfunction} and \eqref{pngjoint-partitionfunction}; second, we have not only accounted for the contributions from the saddle points but also incorporated contributions from their neighborhoods. Our results indicate that, although the conditional likelihood assuming only the lowest-order noise term via saddle-point approximation method is identical to that derived under Gaussian initial conditions (consistent with \cite{Cabass:2019lqx}), these two effects yield irreducible, field-dependent additional contributions. This suggests that, in the context of forward modeling, it is appropriate to introduce these additional terms corresponding to PNG within the conditional likelihood.

In our calculation, we employed the saddle-point expansion method, which is crucial in the computation of decay rates. This method enables us to approximate the path integral by considering both the saddle-point contributions and the contributions from their neighborhoods. When evaluating the latter, it is necessary to consider the issue of negative-mode solutions for the second derivative of the action at the saddle points. Should such solutions exist, we need to systematically deform the original integration contour into the sum of the steepest descent contours in the complexified field space, ensuring a real and convergent result. The application of the steepest descent method in the computation of the likelihoods is analogous to its use in quantum field theory with multiple fields \cite{Garbrecht:2015yza}, as clearly demonstrated in both \cite{Ke:2025uou} and the examples presented in this work. At the same time, we contend that the incorporation of higher-order noise terms is of significant importance, as their inclusion effectively cancels out the UV cutoff dependence inherent in the loop diagrams, analogous to the renormalization procedure in quantum field theory \cite{Schwartz:2014sze}. Therefore, we believe that the corrections to the non-Gaussian conditional likelihood presented in this paper are essential for deriving more stringent constraints on the parameters associated with PNG.

Comparing the expressions for the likelihoods with and without the inclusion of PNG is also of great interest. After the implementation of leading-order PNG, identical additional terms are introduced into the partition functions of both the matter and galaxy fields. However, once an additional source of noise is taken into account, the saddle-point contributions to the matter likelihood and joint likelihood integrals no longer cancel out, thereby justifying the inclusion of corresponding effective terms in simulations. Another noteworthy point is that, even in the absence of cross stochasticities between the matter and galaxy fields, the results of the two likelihoods remain inconsistent around the saddle points. Therefore, we regard our results as both meaningful and of practical relevance.

There remain unresolved issues in our work that cannot be addressed from a purely theoretical perspective. In our calculations, three parameters are suppressed on large scales: the PNG parameter $f_{\rm NL}$, higher-order noise terms, and higher-order expansions around the saddle points. We treat these parameters as perturbative quantities, whose higher-order contributions are progressively suppressed. However, we are currently unable to compare the relative magnitudes of these contributions, although certain parameters can be partially disentangled by performing a rescaling of the momentum. For instance, we may be interested in comparing the contributions around the saddle points with respect to the order of non-Gaussian perturbations, as this enables a more precise extraction of the relevant perturbative terms. These aspects require further investigation through simulations and observational data, and Bayesian forward modeling will, in return, help us strengthen our theoretical framework.

\acknowledgments
JYK thanks Minxing Li for his useful discussions. We thank the anonymous reviewer for his/her professional and invaluable comments and suggestions. We acknowledge the support by the National Science Foundation of China (No. 12147217, 12347163), the China Postdoctoral Science Foundation (No. 2024M761110), and the Natural Science Foundation of Jilin Province, China (No. 20180101228JC).

\bibliographystyle{JHEP}
\bibliography{reference}

\providecommand{\href}[2]{#2}\begingroup\raggedright\begin{thebibliography}{10}

\bibitem{Guth:1980zm}
A.H.~Guth, \emph{{The Inflationary Universe: A Possible Solution to the Horizon and Flatness Problems}}, \href{https://doi.org/10.1103/PhysRevD.23.347}{\emph{Phys. Rev. D} {\bfseries 23} (1981) 347}.

\bibitem{Achucarro:2022qrl}
A.~Ach\'ucarro et~al., \emph{{Inflation: Theory and Observations}},  \href{https://arxiv.org/abs/2203.08128}{{\ttfamily 2203.08128}}.

\bibitem{Linde:1981mu}
A.D.~Linde, \emph{{A New Inflationary Universe Scenario: A Possible Solution of the Horizon, Flatness, Homogeneity, Isotropy and Primordial Monopole Problems}}, \href{https://doi.org/10.1016/0370-2693(82)91219-9}{\emph{Phys. Lett. B} {\bfseries 108} (1982) 389}.

\bibitem{Guth:1982ec}
A.H.~Guth and S.Y.~Pi, \emph{{Fluctuations in the New Inflationary Universe}}, \href{https://doi.org/10.1103/PhysRevLett.49.1110}{\emph{Phys. Rev. Lett.} {\bfseries 49} (1982) 1110}.

\bibitem{Maldacena:2002vr}
J.M.~Maldacena, \emph{{Non-Gaussian features of primordial fluctuations in single field inflationary models}}, \href{https://doi.org/10.1088/1126-6708/2003/05/013}{\emph{JHEP} {\bfseries 05} (2003) 013} [\href{https://arxiv.org/abs/astro-ph/0210603}{{\ttfamily astro-ph/0210603}}].

\bibitem{Cheung:2007st}
C.~Cheung, P.~Creminelli, A.L.~Fitzpatrick, J.~Kaplan and L.~Senatore, \emph{{The Effective Field Theory of Inflation}}, \href{https://doi.org/10.1088/1126-6708/2008/03/014}{\emph{JHEP} {\bfseries 03} (2008) 014} [\href{https://arxiv.org/abs/0709.0293}{{\ttfamily 0709.0293}}].

\bibitem{Lee:2016vti}
H.~Lee, D.~Baumann and G.L.~Pimentel, \emph{{Non-Gaussianity as a Particle Detector}}, \href{https://doi.org/10.1007/JHEP12(2016)040}{\emph{JHEP} {\bfseries 12} (2016) 040} [\href{https://arxiv.org/abs/1607.03735}{{\ttfamily 1607.03735}}].

\bibitem{Arkani-Hamed:2015bza}
N.~Arkani-Hamed and J.~Maldacena, \emph{{Cosmological Collider Physics}},  \href{https://arxiv.org/abs/1503.08043}{{\ttfamily 1503.08043}}.

\bibitem{Dalal:2007cu}
N.~Dalal, O.~Dore, D.~Huterer and A.~Shirokov, \emph{{The imprints of primordial non-gaussianities on large-scale structure: scale dependent bias and abundance of virialized objects}}, \href{https://doi.org/10.1103/PhysRevD.77.123514}{\emph{Phys. Rev. D} {\bfseries 77} (2008) 123514} [\href{https://arxiv.org/abs/0710.4560}{{\ttfamily 0710.4560}}].

\bibitem{Giannantonio:2009ak}
T.~Giannantonio and C.~Porciani, \emph{{Structure formation from non-Gaussian initial conditions: multivariate biasing, statistics, and comparison with N-body simulations}}, \href{https://doi.org/10.1103/PhysRevD.81.063530}{\emph{Phys. Rev. D} {\bfseries 81} (2010) 063530} [\href{https://arxiv.org/abs/0911.0017}{{\ttfamily 0911.0017}}].

\bibitem{Tellarini:2015faa}
M.~Tellarini, A.J.~Ross, G.~Tasinato and D.~Wands, \emph{{Non-local bias in the halo bispectrum with primordial non-Gaussianity}}, \href{https://doi.org/10.1088/1475-7516/2015/07/004}{\emph{JCAP} {\bfseries 07} (2015) 004} [\href{https://arxiv.org/abs/1504.00324}{{\ttfamily 1504.00324}}].

\bibitem{Raccanelli:2015oma}
A.~Raccanelli, M.~Shiraishi, N.~Bartolo, D.~Bertacca, M.~Liguori, S.~Matarrese et~al., \emph{{Future Constraints on Angle-Dependent Non-Gaussianity from Large Radio Surveys}}, \href{https://doi.org/10.1016/j.dark.2016.10.006}{\emph{Phys. Dark Univ.} {\bfseries 15} (2017) 35} [\href{https://arxiv.org/abs/1507.05903}{{\ttfamily 1507.05903}}].

\bibitem{Assassi:2015fma}
V.~Assassi, D.~Baumann and F.~Schmidt, \emph{{Galaxy Bias and Primordial Non-Gaussianity}}, \href{https://doi.org/10.1088/1475-7516/2015/12/043}{\emph{JCAP} {\bfseries 12} (2015) 043} [\href{https://arxiv.org/abs/1510.03723}{{\ttfamily 1510.03723}}].

\bibitem{Wang:2024rdf}
Y.~Wang and P.~He, \emph{{Capturing primordial non-Gaussian signatures in the late Universe by multiscale extrema of the cosmic log-density field}}, \href{https://doi.org/10.1103/PhysRevD.111.L041302}{\emph{Phys. Rev. D} {\bfseries 111} (2025) L041302} [\href{https://arxiv.org/abs/2408.13876}{{\ttfamily 2408.13876}}].

\bibitem{Uhlemann:2017tex}
C.~Uhlemann, E.~Pajer, C.~Pichon, T.~Nishimichi, S.~Codis and F.~Bernardeau, \emph{{Hunting high and low: Disentangling primordial and late-time non-Gaussianity with cosmic densities in spheres}}, \href{https://doi.org/10.1093/mnras/stx2623}{\emph{Mon. Not. Roy. Astron. Soc.} {\bfseries 474} (2018) 2853} [\href{https://arxiv.org/abs/1708.02206}{{\ttfamily 1708.02206}}].

\bibitem{Friedrich:2019byw}
O.~Friedrich, C.~Uhlemann, F.~Villaescusa-Navarro, T.~Baldauf, M.~Manera and T.~Nishimichi, \emph{{Primordial non-Gaussianity without tails \textendash{} how to measure fNL with the bulk of the density PDF}}, \href{https://doi.org/10.1093/mnras/staa2160}{\emph{Mon. Not. Roy. Astron. Soc.} {\bfseries 498} (2020) 464} [\href{https://arxiv.org/abs/1912.06621}{{\ttfamily 1912.06621}}].

\bibitem{DESI:2025qqy}
{\scshape DESI} collaboration, \emph{{Validation of the DESI DR2 Measurements of Baryon Acoustic Oscillations from Galaxies and Quasars}},  \href{https://arxiv.org/abs/2503.14742}{{\ttfamily 2503.14742}}.

\bibitem{Chaussidon:2024sxv}
E.~Chaussidon et~al., \emph{{Blinding scheme for the scale-dependence bias signature of local primordial non-Gaussianity for DESI 2024}}, \href{https://doi.org/10.1088/1475-7516/2025/01/135}{\emph{JCAP} {\bfseries 01} (2025) 135} [\href{https://arxiv.org/abs/2406.00191}{{\ttfamily 2406.00191}}].

\bibitem{DESI:2024wki}
{\scshape DESI} collaboration, \emph{{Validating the galaxy and quasar catalog-level blinding scheme for the DESI 2024 analysis}}, \href{https://doi.org/10.1088/1475-7516/2025/01/128}{\emph{JCAP} {\bfseries 01} (2025) 128} [\href{https://arxiv.org/abs/2404.07282}{{\ttfamily 2404.07282}}].

\bibitem{Andrews:2022nvv}
A.~Andrews, J.~Jasche, G.~Lavaux and F.~Schmidt, \emph{{Bayesian field-level inference of primordial non-Gaussianity using next-generation galaxy surveys}}, \href{https://doi.org/10.1093/mnras/stad432}{\emph{Mon. Not. Roy. Astron. Soc.} {\bfseries 520} (2023) 5746} [\href{https://arxiv.org/abs/2203.08838}{{\ttfamily 2203.08838}}].

\bibitem{Euclid:2024ris}
{\scshape Euclid} collaboration, \emph{{Euclid: Field-level inference of primordial non-Gaussianity and cosmic initial conditions}},  \href{https://arxiv.org/abs/2412.11945}{{\ttfamily 2412.11945}}.

\bibitem{Schmidt:2018bkr}
F.~Schmidt, F.~Elsner, J.~Jasche, N.M.~Nguyen and G.~Lavaux, \emph{{A rigorous EFT-based forward model for large-scale structure}}, \href{https://doi.org/10.1088/1475-7516/2019/01/042}{\emph{JCAP} {\bfseries 01} (2019) 042} [\href{https://arxiv.org/abs/1808.02002}{{\ttfamily 1808.02002}}].

\bibitem{Cabass:2019lqx}
G.~Cabass and F.~Schmidt, \emph{{The EFT Likelihood for Large-Scale Structure}}, \href{https://doi.org/10.1088/1475-7516/2020/04/042}{\emph{JCAP} {\bfseries 04} (2020) 042} [\href{https://arxiv.org/abs/1909.04022}{{\ttfamily 1909.04022}}].

\bibitem{Ke:2025uou}
J.-Y.~Ke, Y.~Wang and P.~He, \emph{{Calculating the EFT likelihood via saddle-point expansion}}, \href{https://doi.org/10.1088/1475-7516/2025/04/064}{\emph{JCAP} {\bfseries 04} (2025) 064} [\href{https://arxiv.org/abs/2502.16082}{{\ttfamily 2502.16082}}].

\bibitem{Voivodic:2025quw}
R.~Voivodic, \emph{{Perturbative Likelihoods for Large-Scale Structure of the Universe}},  \href{https://arxiv.org/abs/2505.23750}{{\ttfamily 2505.23750}}.

\bibitem{Cabass:2020nwf}
G.~Cabass and F.~Schmidt, \emph{{The Likelihood for LSS: Stochasticity of Bias Coefficients at All Orders}}, \href{https://doi.org/10.1088/1475-7516/2020/07/051}{\emph{JCAP} {\bfseries 07} (2020) 051} [\href{https://arxiv.org/abs/2004.00617}{{\ttfamily 2004.00617}}].

\bibitem{Schmidt:2020tao}
F.~Schmidt, \emph{{Sigma-Eight at the Percent Level: The EFT Likelihood in Real Space}}, \href{https://doi.org/10.1088/1475-7516/2021/04/032}{\emph{JCAP} {\bfseries 04} (2021) 032} [\href{https://arxiv.org/abs/2009.14176}{{\ttfamily 2009.14176}}].

\bibitem{Baumann:2010tm}
D.~Baumann, A.~Nicolis, L.~Senatore and M.~Zaldarriaga, \emph{{Cosmological Non-Linearities as an Effective Fluid}}, \href{https://doi.org/10.1088/1475-7516/2012/07/051}{\emph{JCAP} {\bfseries 07} (2012) 051} [\href{https://arxiv.org/abs/1004.2488}{{\ttfamily 1004.2488}}].

\bibitem{Carrasco:2012cv}
J.J.M.~Carrasco, M.P.~Hertzberg and L.~Senatore, \emph{{The Effective Field Theory of Cosmological Large Scale Structures}}, \href{https://doi.org/10.1007/JHEP09(2012)082}{\emph{JHEP} {\bfseries 09} (2012) 082} [\href{https://arxiv.org/abs/1206.2926}{{\ttfamily 1206.2926}}].

\bibitem{Carroll:2013oxa}
S.M.~Carroll, S.~Leichenauer and J.~Pollack, \emph{{Consistent effective theory of long-wavelength cosmological perturbations}}, \href{https://doi.org/10.1103/PhysRevD.90.023518}{\emph{Phys. Rev. D} {\bfseries 90} (2014) 023518} [\href{https://arxiv.org/abs/1310.2920}{{\ttfamily 1310.2920}}].

\bibitem{Porto:2013qua}
R.A.~Porto, L.~Senatore and M.~Zaldarriaga, \emph{{The Lagrangian-space Effective Field Theory of Large Scale Structures}}, \href{https://doi.org/10.1088/1475-7516/2014/05/022}{\emph{JCAP} {\bfseries 05} (2014) 022} [\href{https://arxiv.org/abs/1311.2168}{{\ttfamily 1311.2168}}].

\bibitem{Carrasco:2013mua}
J.J.M.~Carrasco, S.~Foreman, D.~Green and L.~Senatore, \emph{{The Effective Field Theory of Large Scale Structures at Two Loops}}, \href{https://doi.org/10.1088/1475-7516/2014/07/057}{\emph{JCAP} {\bfseries 07} (2014) 057} [\href{https://arxiv.org/abs/1310.0464}{{\ttfamily 1310.0464}}].

\bibitem{Konstandin:2019bay}
T.~Konstandin, R.A.~Porto and H.~Rubira, \emph{{The effective field theory of large scale structure at three loops}}, \href{https://doi.org/10.1088/1475-7516/2019/11/027}{\emph{JCAP} {\bfseries 11} (2019) 027} [\href{https://arxiv.org/abs/1906.00997}{{\ttfamily 1906.00997}}].

\bibitem{Senatore:2014via}
L.~Senatore and M.~Zaldarriaga, \emph{{The IR-resummed Effective Field Theory of Large Scale Structures}}, \href{https://doi.org/10.1088/1475-7516/2015/02/013}{\emph{JCAP} {\bfseries 02} (2015) 013} [\href{https://arxiv.org/abs/1404.5954}{{\ttfamily 1404.5954}}].

\bibitem{Bernardeau:2001qr}
F.~Bernardeau, S.~Colombi, E.~Gaztanaga and R.~Scoccimarro, \emph{{Large scale structure of the universe and cosmological perturbation theory}}, \href{https://doi.org/10.1016/S0370-1573(02)00135-7}{\emph{Phys. Rept.} {\bfseries 367} (2002) 1} [\href{https://arxiv.org/abs/astro-ph/0112551}{{\ttfamily astro-ph/0112551}}].

\bibitem{Blas:2015qsi}
D.~Blas, M.~Garny, M.M.~Ivanov and S.~Sibiryakov, \emph{{Time-Sliced Perturbation Theory for Large Scale Structure I: General Formalism}}, \href{https://doi.org/10.1088/1475-7516/2016/07/052}{\emph{JCAP} {\bfseries 07} (2016) 052} [\href{https://arxiv.org/abs/1512.05807}{{\ttfamily 1512.05807}}].

\bibitem{Blas:2016sfa}
D.~Blas, M.~Garny, M.M.~Ivanov and S.~Sibiryakov, \emph{{Time-Sliced Perturbation Theory II: Baryon Acoustic Oscillations and Infrared Resummation}}, \href{https://doi.org/10.1088/1475-7516/2016/07/028}{\emph{JCAP} {\bfseries 07} (2016) 028} [\href{https://arxiv.org/abs/1605.02149}{{\ttfamily 1605.02149}}].

\bibitem{Vasudevan:2019ewf}
A.~Vasudevan, M.M.~Ivanov, S.~Sibiryakov and J.~Lesgourgues, \emph{{Time-sliced perturbation theory with primordial non-Gaussianity and effects of large bulk flows on inflationary oscillating features}}, \href{https://doi.org/10.1088/1475-7516/2019/09/037}{\emph{JCAP} {\bfseries 09} (2019) 037} [\href{https://arxiv.org/abs/1906.08697}{{\ttfamily 1906.08697}}].

\bibitem{Rubira:2024tea}
H.~Rubira and F.~Schmidt, \emph{{The renormalization group for large-scale structure: origin of galaxy stochasticity}}, \href{https://doi.org/10.1088/1475-7516/2024/10/092}{\emph{JCAP} {\bfseries 10} (2024) 092} [\href{https://arxiv.org/abs/2404.16929}{{\ttfamily 2404.16929}}].

\bibitem{Nikolis:2024kbx}
C.~Nikolis, H.~Rubira and F.~Schmidt, \emph{{The renormalization group for large-scale structure: primordial non-Gaussianities}}, \href{https://doi.org/10.1088/1475-7516/2024/08/017}{\emph{JCAP} {\bfseries 08} (2024) 017} [\href{https://arxiv.org/abs/2405.21002}{{\ttfamily 2405.21002}}].

\bibitem{Coleman:1977py}
S.R.~Coleman, \emph{{The Fate of the False Vacuum. 1. Semiclassical Theory}}, \href{https://doi.org/10.1103/PhysRevD.16.1248}{\emph{Phys. Rev. D} {\bfseries 15} (1977) 2929}.

\bibitem{Callan:1977pt}
C.G.~Callan, Jr. and S.R.~Coleman, \emph{{The Fate of the False Vacuum. 2. First Quantum Corrections}}, \href{https://doi.org/10.1103/PhysRevD.16.1762}{\emph{Phys. Rev. D} {\bfseries 16} (1977) 1762}.

\bibitem{Schmidt:2010gw}
F.~Schmidt and M.~Kamionkowski, \emph{{Halo Clustering with Non-Local Non-Gaussianity}}, \href{https://doi.org/10.1103/PhysRevD.82.103002}{\emph{Phys. Rev. D} {\bfseries 82} (2010) 103002} [\href{https://arxiv.org/abs/1008.0638}{{\ttfamily 1008.0638}}].

\bibitem{Assassi:2015jqa}
V.~Assassi, D.~Baumann, E.~Pajer, Y.~Welling and D.~van~der Woude, \emph{{Effective theory of large-scale structure with primordial non-Gaussianity}}, \href{https://doi.org/10.1088/1475-7516/2015/11/024}{\emph{JCAP} {\bfseries 11} (2015) 024} [\href{https://arxiv.org/abs/1505.06668}{{\ttfamily 1505.06668}}].

\bibitem{Schwartz:2014sze}
M.D.~Schwartz, \emph{{Quantum Field Theory and the Standard Model}}, Cambridge University Press (3, 2014).

\bibitem{Ai:2019fri}
W.-Y.~Ai, B.~Garbrecht and C.~Tamarit, \emph{{Functional methods for false vacuum decay in real time}}, \href{https://doi.org/10.1007/JHEP12(2019)095}{\emph{JHEP} {\bfseries 12} (2019) 095} [\href{https://arxiv.org/abs/1905.04236}{{\ttfamily 1905.04236}}].

\bibitem{Andreassen:2016cvx}
A.~Andreassen, D.~Farhi, W.~Frost and M.D.~Schwartz, \emph{{Precision decay rate calculations in quantum field theory}}, \href{https://doi.org/10.1103/PhysRevD.95.085011}{\emph{Phys. Rev. D} {\bfseries 95} (2017) 085011} [\href{https://arxiv.org/abs/1604.06090}{{\ttfamily 1604.06090}}].

\bibitem{Witten:2010cx}
E.~Witten, \emph{{Analytic Continuation Of Chern-Simons Theory}}, {\emph{AMS/IP Stud. Adv. Math.} {\bfseries 50} (2011) 347} [\href{https://arxiv.org/abs/1001.2933}{{\ttfamily 1001.2933}}].

\bibitem{Garbrecht:2015oea}
B.~Garbrecht and P.~Millington, \emph{{Green\textquoteright{}s function method for handling radiative effects on false vacuum decay}}, \href{https://doi.org/10.1103/PhysRevD.91.105021}{\emph{Phys. Rev. D} {\bfseries 91} (2015) 105021} [\href{https://arxiv.org/abs/1501.07466}{{\ttfamily 1501.07466}}].

\bibitem{Quiros:1999jp}
M.~Quiros, \emph{{Finite temperature field theory and phase transitions}},  in \emph{{ICTP Summer School in High-Energy Physics and Cosmology}}, pp.~187--259, 1, 1999 [\href{https://arxiv.org/abs/hep-ph/9901312}{{\ttfamily hep-ph/9901312}}].

\bibitem{Moreno:1998bq}
J.M.~Moreno, M.~Quiros and M.~Seco, \emph{{Bubbles in the supersymmetric standard model}}, \href{https://doi.org/10.1016/S0550-3213(98)00283-1}{\emph{Nucl. Phys. B} {\bfseries 526} (1998) 489} [\href{https://arxiv.org/abs/hep-ph/9801272}{{\ttfamily hep-ph/9801272}}].

\bibitem{Ke:2024lel}
J.~Ke, M.~Li and P.~He, \emph{{Principle of multi-critical-points in the ALP-Higgs model and the corresponding phase transition}}, \href{https://doi.org/10.1016/j.physletb.2024.138546}{\emph{Phys. Lett. B} {\bfseries 850} (2024) 138546} [\href{https://arxiv.org/abs/2402.12085}{{\ttfamily 2402.12085}}].

\bibitem{Chen:2009zp}
X.~Chen and Y.~Wang, \emph{{Quasi-Single Field Inflation and Non-Gaussianities}}, \href{https://doi.org/10.1088/1475-7516/2010/04/027}{\emph{JCAP} {\bfseries 04} (2010) 027} [\href{https://arxiv.org/abs/0911.3380}{{\ttfamily 0911.3380}}].

\bibitem{Komatsu:2001rj}
E.~Komatsu and D.N.~Spergel, \emph{{Acoustic signatures in the primary microwave background bispectrum}}, \href{https://doi.org/10.1103/PhysRevD.63.063002}{\emph{Phys. Rev. D} {\bfseries 63} (2001) 063002} [\href{https://arxiv.org/abs/astro-ph/0005036}{{\ttfamily astro-ph/0005036}}].

\bibitem{Alishahiha:2004eh}
M.~Alishahiha, E.~Silverstein and D.~Tong, \emph{{DBI in the sky}}, \href{https://doi.org/10.1103/PhysRevD.70.123505}{\emph{Phys. Rev. D} {\bfseries 70} (2004) 123505} [\href{https://arxiv.org/abs/hep-th/0404084}{{\ttfamily hep-th/0404084}}].

\bibitem{Green:2013rd}
D.~Green, M.~Lewandowski, L.~Senatore, E.~Silverstein and M.~Zaldarriaga, \emph{{Anomalous Dimensions and Non-Gaussianity}}, \href{https://doi.org/10.1007/JHEP10(2013)171}{\emph{JHEP} {\bfseries 10} (2013) 171} [\href{https://arxiv.org/abs/1301.2630}{{\ttfamily 1301.2630}}].

\bibitem{Flauger:2016idt}
R.~Flauger, M.~Mirbabayi, L.~Senatore and E.~Silverstein, \emph{{Productive Interactions: heavy particles and non-Gaussianity}}, \href{https://doi.org/10.1088/1475-7516/2017/10/058}{\emph{JCAP} {\bfseries 10} (2017) 058} [\href{https://arxiv.org/abs/1606.00513}{{\ttfamily 1606.00513}}].

\bibitem{Rubira:2023vzw}
H.~Rubira and F.~Schmidt, \emph{{Galaxy bias renormalization group}}, \href{https://doi.org/10.1088/1475-7516/2024/01/031}{\emph{JCAP} {\bfseries 01} (2024) 031} [\href{https://arxiv.org/abs/2307.15031}{{\ttfamily 2307.15031}}].

\bibitem{Planck:2019kim}
{\scshape Planck} collaboration, \emph{{Planck 2018 results. IX. Constraints on primordial non-Gaussianity}}, \href{https://doi.org/10.1051/0004-6361/201935891}{\emph{Astron. Astrophys.} {\bfseries 641} (2020) A9} [\href{https://arxiv.org/abs/1905.05697}{{\ttfamily 1905.05697}}].

\bibitem{Weinberg:1987vp}
E.J.~Weinberg and A.-q.~Wu, \emph{{UNDERSTANDING COMPLEX PERTURBATIVE EFFECTIVE POTENTIALS}}, \href{https://doi.org/10.1103/PhysRevD.36.2474}{\emph{Phys. Rev. D} {\bfseries 36} (1987) 2474}.

\bibitem{Garbrecht:2015yza}
B.~Garbrecht and P.~Millington, \emph{{Self-consistent solitons for vacuum decay in radiatively generated potentials}}, \href{https://doi.org/10.1103/PhysRevD.92.125022}{\emph{Phys. Rev. D} {\bfseries 92} (2015) 125022} [\href{https://arxiv.org/abs/1509.08480}{{\ttfamily 1509.08480}}].

\bibitem{Jasche:2012kq}
J.~Jasche and B.D.~Wandelt, \emph{{Bayesian physical reconstruction of initial conditions from large scale structure surveys}}, \href{https://doi.org/10.1093/mnras/stt449}{\emph{Mon. Not. Roy. Astron. Soc.} {\bfseries 432} (2013) 894} [\href{https://arxiv.org/abs/1203.3639}{{\ttfamily 1203.3639}}].

\bibitem{Wang:2014hia}
H.~Wang, H.J.~Mo, X.~Yang, Y.P.~Jing and W.P.~Lin, \emph{{ELUCID - Exploring the Local Universe with reConstructed Initial Density field I: Hamiltonian Markov Chain Monte Carlo Method with Particle Mesh Dynamics}}, \href{https://doi.org/10.1088/0004-637X/794/1/94}{\emph{Astrophys. J.} {\bfseries 794} (2014) 94} [\href{https://arxiv.org/abs/1407.3451}{{\ttfamily 1407.3451}}].

\bibitem{Jasche:2018oym}
J.~Jasche and G.~Lavaux, \emph{{Physical Bayesian modelling of the non-linear matter distribution: new insights into the Nearby Universe}}, \href{https://doi.org/10.1051/0004-6361/201833710}{\emph{Astron. Astrophys.} {\bfseries 625} (2019) A64} [\href{https://arxiv.org/abs/1806.11117}{{\ttfamily 1806.11117}}].

\bibitem{Kitaura:2019ber}
F.-S.~Kitaura, M.~Ata, S.A.~Rodriguez-Torres, M.~Hernandez-Sanchez, A.~Balaguera-Antolinez and G.~Yepes, \emph{{COSMIC BIRTH: Efficient Bayesian Inference of the Evolving Cosmic Web from Galaxy Surveys}}, \href{https://doi.org/10.1093/mnras/staa3774}{\emph{Mon. Not. Roy. Astron. Soc.} {\bfseries 502} (2021) 3456} [\href{https://arxiv.org/abs/1911.00284}{{\ttfamily 1911.00284}}].

\bibitem{Bos:2018rpw}
E.G.P.~Bos, F.-S.~Kitaura and R.~van~de Weygaert, \emph{{Bayesian cosmic density field inference from redshift space dark matter maps}}, \href{https://doi.org/10.1093/mnras/stz1864}{\emph{Mon. Not. Roy. Astron. Soc.} {\bfseries 488} (2019) 2573} [\href{https://arxiv.org/abs/1810.05189}{{\ttfamily 1810.05189}}].

\bibitem{Baumann:2009ds}
D.~Baumann, \emph{{Inflation}},  in \emph{{Theoretical Advanced Study Institute in Elementary Particle Physics}: {Physics of the Large and the Small}}, pp.~523--686, 2011, \href{https://doi.org/10.1142/9789814327183_0010}{DOI} [\href{https://arxiv.org/abs/0907.5424}{{\ttfamily 0907.5424}}].

\end{thebibliography}\endgroup

\end{document}